\documentclass[12pt, letterpaper]{extarticle}

\usepackage{xcolor}
\usepackage{hyperref}
\usepackage{makecell}
\usepackage{url}
\usepackage{adjustbox}
\usepackage[flushleft]{threeparttable}
\usepackage{booktabs}
\usepackage{graphics}
\usepackage{helvet}
\usepackage{times}
\usepackage{comment}
\usepackage{placeins}
\usepackage{multirow}
\usepackage{rotating}
\usepackage{mathtools}
\usepackage{subfig}
\usepackage{arydshln}
\usepackage{longtable}
\usepackage{float}
\usepackage{multicol}
\usepackage{supertabular}
\usepackage{booktabs}
\usepackage{longtable}
\usepackage[framemethod=TikZ]{mdframed}
\usepackage{fullpage}
\usepackage[switch]{lineno}
\usepackage{amsmath}
\usepackage{amssymb}
\usepackage{rotating}
\usepackage{array}
\usepackage{mathtools}
\usepackage[ruled]{algorithm2e}
\usepackage{algorithmic}
\usepackage{bm}
\usepackage{breqn}
\usepackage{comment}
\usepackage{enumitem}
\usepackage{graphics}
\usepackage{graphicx}
\usepackage{latexsym}
\usepackage{mathrsfs}
\usepackage{morefloats}
\usepackage{nicefrac}
\usepackage{authblk}
\usepackage{ulem}
\usepackage{siunitx}
\usepackage{sidecap}
\usepackage{multirow}
\usepackage{xurl}
\usepackage{lscape}

\definecolor{blue}{rgb}{0.0, 0.0, 1.0}
\definecolor{normgreen}{rgb}{0,0.6753,0.3241}

\newcommand{\minus}{\scalebox{0.75}[1.0]{$-$}}

\title{{Current policies governing editorial conflicts of interest are ineffective}}

\author[1,2]{Fengyuan Liu}
\author[3*]{Bedoor AlShebli}
\author[1*]{Talal Rahwan}

\affil[1]{\normalsize Computer Science, Science Division, New York University Abu Dhabi, UAE}

\affil[2]{\normalsize Courant Institute of Mathematical Sciences, New York University, New York, NY 10012, USA}

\affil[3]{\normalsize Social Science Division, New York University Abu Dhabi, UAE}

\affil[*]{\footnotesize Corresponding author. E-mail:\ talal.rahwan@nyu.edu, bedoor@nyu.edu}

\renewcommand{\bf}{}

\date{}

\begin{document}
\maketitle

\baselineskip22pt


\section*{Abstract}
\begin{quote}
\textbf{\small
{
Research-active editors face a potential conflict of interest (COI) when handling submissions from authors who share the same affiliation or those who recently collaborated with the editor. Since perception of COIs arising from such editor-author associations may erode trust in science, some policies recommend, and others demand, recusal in such incidents. However, the effectiveness of such measures is unknown to date. To fill this gap, we analyze half a million papers from six publishers who specify the handling editor of each paper. We find numerous papers with editor-author associations, and demonstrate that such papers tend to be accepted faster. A quasi-experimental design exploiting policy changes at PNAS and PLOS reveals the limited effectiveness of current COI policies. A network neural embedding model reveals that requiring editors with potential COIs to recuse may compromise the suitability of the handling editor. Finally, an online survey experiment demonstrates that such COIs influence trust in the paper's finding, but public disclosure eliminates this effect.
}
}
\end{quote}


\section*{Introduction}
Academic editors play a crucial role in the scientific community as gatekeepers of {scientific publishing}~\cite{siler2015measuring}.
Apart from very few exceptions (e.g., professional editors of \textit{Cell}, \textit{Nature} and \textit{Science}), the vast majority of those who serve as editors do so as a community service while focusing on their primary role as research-active scientists. The dual roles that editors may have could lead to conflicts of interest (COIs), i.e., situations where the private interests of editors could influence, or could be perceived to influence, objective assessment of academic papers.


In this study, we focus on a particular type of COI that all research-active editors could face---non-financial COI due to personal connections. According to the Council of Science Editors, a COI arises when an editor handles a submission (co-)authored by a colleague (i.e., someone affiliated with the same institution) or by a person with whom they collaborated or co-authored recently~\cite{CSE}.
{In such situations, editors with a COI are expected to delegate the decision-making to other editors~\cite{CSE}. Despite policy efforts, the degree to which such policies are being enforced remains unclear. Moreover,} current COI governance might not fully reflect the complex situations in which scientific research takes place. On paper, such policies make sweeping statements that cover all editor-author associations. In practice, as we will later show in our analysis, various factors complicate the implications of such policies.

To date, very few quantitative studies focus on policies governing editors' COI. Existing studies often focus on understanding the prevalence of such policies, rather than their impact, and are restricted to medical journals~\cite{haivas2004editors,bosch2013financial,faggion2021watching,smith2012accessibility}. Here, to understand the {interplay between such policies and how editors handle COIs}, we compile a dataset of half a million papers along with their handling editors from six different publishers, namely, Frontiers, Hindawi, IEEE, MDPI, PLOS, and PNAS. Note that due to the limited availability of data related to the handling editors of academic publications, we only examine publishers that openly share such data, bearing in mind that this is not necessarily representative of all academic publishers. 
{Using this dataset, we find that editor-author associations are not uncommon, even among publishers with strict recusal policies. Leveraging a quasi-experimental design, we further reveal that recusal policies have a limited effect on the rate of papers with such associations. One possible explanation could be that journals prioritize fit of editorial expertise over absence of COIs. To explore this possibility, we quantify the expertise of handling editors using a network embedding model. Based on this, we estimate that 30\% of papers would have been handled by less suitable editors if COI policies had been enforced. This finding highlights a potential integrity-suitability tradeoff when determining the handling editor. Finally, we propose public disclosure as an alternative form of governance, and demonstrate its effect on public trust in science using an online survey experiment. Taken together, this study} addresses the need to inform the development and implementation of COI policies~\cite{plos2008making}.

{\section*{An overview of policies governing editor-author associations}
}
Before delving into quantitative analysis, we provide a qualitative overview of the policies governing editors' COI, gathered from four major organizations of academic editors, namely, the Committee on Publication Ethics (COPE)~\cite{COPE_COI}, the Council of Science Editors (CSE)~\cite{CSE}, the International Committee of Medical Journal Editors (ICMJE)~\cite{ICMJE}, and the World Association of Medical Editors (WAME)~\cite{WAME2009COI}. Additionally, we examine the COI policies of the six academic publishers analyzed in this study~\cite{FrontiersPolicy,HindawiPolicy,IEEEpolicy,MDPIpolicy,PLOSpolicy,PNASpolicy}; see Supplementary Table~1 for the relevant clauses extracted from these policies, and Table~\ref{tab:policy} and~\ref{tab:publishers} for a summary.

{Formally, COI arises when ``there is a divergence between an individual’s private interests (competing interests) and his or her responsibilities to scientific and publishing activities such that a reasonable observer might wonder if the individual’s behavior or judgment was motivated by considerations of his or her competing interests.''~\cite{WAME2009COI} This definition emphasizes that even the perception of COI can undermine trust in the editorial process, even when lacking actual COIs. Accordingly, although editor-author associations may not always introduce actual COIs, from a governance perspective, we do not distinguish between potential and actual COIs.}

Overall, there is no consensus on types of editor-author associations that would constitute a COI. However, several types of associations are commonly mentioned across policy documents. Among these, two of the most frequently mentioned ones are recent collaborations and shared institutional affiliations between the editor and author---both of which are covered in our study. Additional relationships commonly identified as potential sources of COI include familial ties~\cite{WAME2009COI,FrontiersPolicy,MDPIpolicy,PNASpolicy}, direct competition~\cite{CSE,WAME2009COI}, mentor-mentee relationships~\cite{COPE_COI,MDPIpolicy}, and joint grant partnerships~\cite{PLOSpolicy}.

We start by reviewing policies governing editor-author associations arising from shared affiliations. This is explicitly identified as a source of COI by three academic organizations (COPE, CSE, and WAME) as well as four academic publishers (Frontiers, Hindawi, MDPI, and PLOS). Among them, CSE adopts the most lenient policy, since it only considers being affiliated with the same department, not institution, as a source of COI. Similarly, WAME's policy is relatively relaxed because, {in addition to affiliation with the same department}, it considers affiliation with the same institution as a source of COI only when the affiliation is ``small'', although it does not provide any guidelines on what should be considered ``small.'' {Such ambiguity further extends to independent institutes (e.g., the Knight First Amendment Institute), academic colleges (e.g., Barnard College), hospital departments (e.g., NYU Langone Health), etc., which are part of bigger universities but tend to operate independently as ``smaller'' institutions on their own.} On the contrary, Hindawi adopts the strictest policy, whereby COI exists not only when the editor and author share the same affiliation currently, but also in the recent past.

Next, regarding editor-author association due to recent collaborations, two academic organizations (COPE and CSE) and five academic publishers (all except IEEE) explicitly address this type of editor-author association, but they disagree as to what constitutes a ``recent collaboration''. More specifically, according to PNAS, a collaboration is considered recent if the difference between the date on which it took place and the date of submission ($\Delta$) does not exceed 48 months (i.e., if $\Delta \leq 48\textnormal{m}$)~\cite{PNASpolicy}; see Methods for a more formal definition. In contrast, Frontiers, MPDI, and PLOS consider the threshold to be {2 years, 3 years, and 5 years}, respectively~\cite{FrontiersPolicy,MDPIpolicy,PLOSpolicy}.
As for the two remaining publishers in our dataset, Hindawi does not provide an explicit definition of what counts as a recent collaboration~\cite{HindawiPolicy}, while IEEE does not explicitly mention editor-author collaboration in its policies~\cite{IEEEpolicy,IEEEpubPolicy}.

{Editors' COI is managed in two primary ways: disclosure and recusal. Most policy documents reviewed in this paper recommend or mandate disclosure of editors' COI, but in practice such disclosures are never publicly released along with the paper in question. Most policies further recommend or mandate that editors recuse themselves from handling submissions with COI. The stringency of these recusal policies, however, varies across publishers, which we classify into three tiers based on the language used therein; see the ``Interpretation of policies'' subsection of Methods for more details. Specifically, Frontiers and PLOS fall in the first (least stringent) tier: Frontiers simply asks editors to consider their COI before handling a paper, while PLOS suggests that editors may recuse themselves when a COI exists. At the opposite end, PNAS falls in the third (most stringent) tier, forbidding editors from handling papers with COI. All other publishers, as well as all academic associations (i.e., WAME, CSE, ICMJE, and COPE), occupy the middle ground, recommending against editors' handling papers with COI, but do not prohibit such interaction}.

\vspace{0.5cm}
\section*{Results}
{
\subsection*{The percentage of papers with editor-author associations}}
{We start by documenting the frequencies of editor-author associations observed in our dataset.} As can be seen in Fig.~\ref{coi_rate}a--c, the percentage of papers with an editor-author association varies across publishers. PNAS tops the chart with 10.5\% of papers having recent editor-author collaboration, 6.9\% of papers where the editor and an author share the same affiliation, and 15.3\% of papers having either type of editor-author association; see Supplementary Table~2 for detailed statistics.
As for the journals, the ones with a high percentage of papers with editor-author association are PLOS Medicine ({24}\%), Frontiers in Pediatrics ({16}\%), PNAS ({15}\%), Journal of Fungi ({15}\%), PLOS Neglected Tropical Diseases ({15}\%), and Frontiers in Neuroinformatics ({14}\%); all these journals are ranked in the first quartile (Q1) in their respective disciplines. In fact, 16 out of the 20 journals with the highest {percentage} are ranked Q1; see Supplementary Table~3. Overall, nearly 6\% of journals have a {percentage} $\geq$ 10\%, and over half of them have a {percentage} $\geq$ 2\%.


{Next, we focus on special issues, since they may arguably have a higher percentage of papers with editor-author associations.
This may stem from a less rigorous peer review process~\cite{science2023stripped,else2021scammers}, as evidenced by the disproportionately large number of retractions from such issues~\cite{mills2024special,van2023more}. Another contributing factor may be the limited pool of scholars with suitable expertise in highly specialized fields~\cite{pfeffer2018queer}, since special issues tend to ``focus on a specific area of research~\cite{else2021tortured}.''} Motivated by this observation, we next compare special issues with ``normal'' ones in terms of the percentages of papers with editor-author associations. As can be seen in Supplementary Table~4, the percentages of papers with editor-author association in special issues are indeed greater than those in normal editions. More specifically, the percentage in Hindawi drops from 6.68\% in special issues to merely 1\% in normal issues, the percentage drops from 9.95\% to 4.35\% in Frontiers, and from 2.77\% to 2.01\% in MDPI.

\subsection*{The characteristics of papers with editor-author associations}
Next, we determine whether the percentages of papers with editor-author association varies with the editor's characteristics. We find that the percentage increases with the affiliation rank of handling editors, and varies across disciplines. Specifically, the {percentages} in Biology, Chemistry, and Medicine reach over 6\%, while Business has the lowest rate of about 2.7\% (Fig.~\ref{coi_rate}{e}). In Supplementary Figure~2, we find broadly similar patterns using a logistic regression model that controls for the affiliation, and discipline of the handling editor, in addition to journal age (number of years since the birth of the journal), editor age (number of years since the first time that the editor handles a paper in this journal), and the number of authors of the paper in question.

We further examine the nature of these editor-author associations, and find that they are not evenly distributed across different attributes.
Firstly, in over 65\% of papers with editor-author association, the editor was associated with either the first or the last author (Fig.~\ref{coi_rate}f). Secondly, in 53\% of papers with recent editor-author collaborations, the author have collaborated more than once with the editor during the 48 months prior to submitting the paper in question (Fig.~\ref{coi_rate}g). Thirdly, in over 47\% of papers with recent editor-author collaborations, the most recent collaboration took place less than a year prior to the papers' submission (Fig.~\ref{coi_rate}h).

Finally, we find that in over 81\% of papers with recent editor-author collaborations, the prior collaborations happened in teams involving no more than 20 authors (Fig.~\ref{coi_rate}i). Note that different disciplines have different distributions of team size. For example, the 75-th percentile value of team size is 4 in Philosophy and 7 in Medicine, and the 95-th percentile value of team size is 7 in Philosophy and 12 in Medicine (Supplementary Table~5). {These differences reflect disciplinary variations in terms of: research pace~\cite{becher1994significance}, strategy of inquiry~\cite{jaffe2014social}, and resource requirements~\cite{wray2022epistemic}. As a result, the natural sciences tend to be more collaborative and competitive, whereas the humanities and many social sciences are characterized by smaller research teams \cite{lariviere2006canadian}. On average,} disciplines with larger teams tend to have a higher percentage of papers featuring recent editor-author collaborations (Supplementary Figure~3). Notably, Medicine and Chemistry exhibit the highest percentages, exceeding 4.5\%, while Engineering and Business show the lowest, around 2\% (Supplementary Table~6). Overall, 54\% of papers with recent editor-author collaborations falls in either Biology, Medicine, or both. These two fields are also the most productive, accounting for 47\% of the papers in our dataset. In each discipline, if we disregard all prior collaborations whose team size falls above the 95-th percentile in that discipline, between 56.41\% to 78.46\% of editor-author collaborations would still persist, suggesting that at least half of recent editor-author collaborations took place in small to medium-sized teams. If one were to further restrict their attention only to teams that fall under the 75th percentile value, between one-fifth (in Business) to half (in Engineering) of recent editor-author collaborations persist.

\subsection*{The acceptance delay of papers with recent editor-author collaborations}
We now investigate whether they differ from other papers in terms of the time spent between submission and acceptance. We further account for the temporal and cross-sectional variation in acceptance delay by calculating relative acceptance delay (RAD); see Methods for more details. 
As can be seen in Fig.~\ref{rad}a, papers with {recent editor-author collaboration} have a shorter RAD than those without. In particular, the RAD of papers without {recent editor-author collaboration} is normally distributed around 0 (mean: 1.56, standard deviation: 64.74), whereas the RAD of papers with {recent editor-author collaboration} has two modes, with one of them roughly around 0 and the other around $\minus$43. Such a bimodal distribution suggests that papers with {recent editor-author collaboration} consist of two distinct subpopulations, with one being handled at the usual pace (i.e., just like papers without recent editor-author collaboration), and the other being handled faster.

{Relying on observational data only, we are unable to establish the causal relationship between editor-author collaboration and faster acceptance. There could be many explanations for this phenomenon. One such explanation is that papers involving editor–author associations may have higher potential impact~\cite{brogaard2014networks}, leading editors to prioritize them in an effort to maximize the expected citation count of accepted papers~\cite{card2020editors}. Nevertheless, we find that the differences in citation counts do not explain the differences in acceptance delays. More specifically, as shown in Supplementary Figure~4, papers with editor-author associations are accepted faster, even when accounting for citation count.
Finally, we account for potential disciplinary variation in acceptance delay. While RAD already captures within-journal variations in peer review timeline, there could still be disciplinary variations within multidisciplinary journals such as PNAS and PLOS One. To this end, we repeat the same analysis while explicitly controlling for discipline fixed-effects, revealing that the association between prior editor–author collaboration and shorter RAD persists (Supplementary Table~7).
}

Next, we explore whether papers whose authors have a stronger relationship with the editor are accepted faster than those with a weaker relationship. We first look at how RAD is related to author positions.
Here, when referring to a paper, $p$, we will add an asterisk superscript ($p^*$) to indicate that the handling editor has collaborated with the first or last author of $p$. Furthermore, we will add a dagger superscript ($p^{\dagger}$) to indicate that an author of $p$ is the first or last author in a prior collaboration with the editor. Finally, two daggers ($p^{\dagger\dagger}$) would indicate that the editor and an author of $p$ were the first and last authors in a prior collaboration. Note that these notations can be used simultaneously, e.g., we could write $p^{*\dagger}$, or $p^{*\dagger\dagger}$, depending on the paper type. {The absence of any superscripts indicates that the editor has a prior collaboration with a middle author of $p$, and that both of them where middle authors in their prior collaboration.}
Arguably, the link between the submission and the editor tends to be stronger in $p^*$ than in $p$, and also tends to be stronger in $p^{\dagger}$ and $p^{\dagger\dagger}$ than in $p$. This argument is motivated the observation that, compared to middle authors, first and last authors are significantly more likely to be corresponding authors, as well as having broader involvement in research activities~\cite{sauermann2017authorship}, which means they play more important roles in a research paper as well as more likely to liaise with the handling editor. As can be seen in Fig.~\ref{rad}b, $p^{*}$ is handled faster than $p$. Similarly, $p^{*\dagger}$ is handled faster than $p^{\dagger}$, and $p^{*\dagger\dagger}$ is handled faster than $p^{\dagger\dagger}$. Finally, both $p^{*\dagger}$ and $p^{*\dagger\dagger}$ are handled faster than $p^{*}$.

We then calculate the percentage of authors who have recently collaborated with the editor. As can be seen in {Fig.~\ref{rad}c}, the greater the percentage, the shorter the RAD. {A third} way to examine relationship strength between an author and an editor is to consider the team size of their prior collaboration. Intuitively, if an author has collaborated with the editor as part of a smaller team (e.g., involving, say, three members), then the author-editor relationship is likely to be stronger than if they were part of a larger team (e.g., involving, say, 20 members). As can be seen in {Fig.~\ref{rad}d}, the smaller the team size, the shorter the RAD. These findings suggest that authors with a stronger connection to the editor experience shorter delays between the submission and acceptance of their manuscripts. More importantly, Fig.~\ref{rad}d suggests that the shorter RAD observed earlier (i.e., the one enjoyed by papers with recent editor-author collaborations) diminishes when the collaborations involve large teams.

To identify the team size beyond which the effect disappears, we grouped papers based on the team size of recent editor-author collaborations, and then applied a one-sample t-test to examine whether the RAD of each group of papers are significantly different from zero, while ensuring similar statistical power in each group. As can be seen in Supplementary Figure~5, when recent editor-author collaborations involve teams with fewer than 20 co-authors, the corresponding papers tend to be accepted faster than a typical paper published in the same journal in the same year. However, this difference in RAD disappears when recent editor-author collaboration involves teams of 20 or more co-authors. {
Since different disciplines tend to have different team size distributions (Supplementary Table~5), we next repeat the same analysis but for each discipline separately. As can be seen in Supplementary Figure~6, in the Social Sciences, Computer Science, Engineering, and Math, papers with prior editor-author collaborations are accepted faster only when the prior collaboration happens in teams of no more than six co-authors. This number is bigger in Physics (11), and even bigger in Biology, Chemistry, and Medicine (21).

Taken together, these findings raise the possibility of favoritism in editorial decisions, as editors who have previously collaborated with the authors tend to accept their papers faster, especially when the prior collaboration involves a relatively small team. These results suggest that policies should consider team size when determining whether a prior editor-author collaboration introduces a potential COI, possibly cases involving teams larger than a certain threshold. Ideally, such a threshold should vary across disciplines to reflect variations in publishing norms. That said, it is important to emphasize that our findings are based on purely observational data, and thus the observed pattern could still be correlational rather than causal.}


\vspace{0.5cm}
{
\subsection*{The effect of recusal policies}
}

{As we have seen in Table~\ref{tab:policy}, many publishers have adopted recusal as a way to curb papers with editor-author associations, but} it remains unclear whether these policies have any effect at all. In other words, what percentage of papers with editor-author associations do we expect to see in the absence of these policies?

Answering this counterfactual question based on observational data alone is challenging, since we cannot observe a parallel universe in which the policies were never introduced. Nevertheless, we are able to estimate the policies' effect by leveraging three quasi-experiments whereby certain changes were introduced to the COI policies of PNAS and PLOS at different points in time. Based on this, we set out to compare editors' behaviour before vs.~after the changes were introduced.

More specifically, the first policy change that we analyze (Case~1) took place in July 2011, when PNAS introduced a COI policy, prohibiting editors from handling submissions by authors with whom they collaborated during the past 24 months. Importantly, no such policy existed prior to that date. The second policy change that we analyze (Case~2) took place in January 2014, when PNAS updated its policy by modifying its definition of ``recent collaboration'' from the past 24 months to the past 48 months. The third and final policy change (Case~3) took place in May 2015, when PLOS introduced a COI policy, {recommending against} editors from handling submissions by authors with whom they collaborated during the past 60 months; no such policy existed in PLOS prior to that date. See supplementary materials for more details regarding these policy changes. 

In all three cases of policy change, the scope of editor-author collaborations considered as a source of potential COI was broadened. To put it differently, certain behavior that was considered acceptable by the publisher became prohibited as per the new policy. {Despite this common attribute, three cases of policy change have some subtle distinctions. Unlike Case~1 and Case~2, where PNAS states that ``recent collaborators ... must be excluded as editors'', Case~3 (PLOS) adopts less strict policy by only suggesting that editors may recuse themselves if necessary (see Supplementary Note 1 for policy text). Additionally, unlike Case~1 and Case~3, where COI policies were introduced for the first time in their respective publisher, Case 2 was an update of an existing policy in PNAS.}

We start by visualizing the percentage of papers with {recent editor-author collaborations} that were submitted around the month in which the policy change took place (Fig.~\ref{policy_change}a to \ref{policy_change}c). To estimate the effect of the policy change, we use a regression discontinuity in time (RDiT) design, a method commonly used in the Social Sciences to study the treatment effect in quasi-experiments~\cite{imbens2008regression,anderson2014subways,reny2021opinion}. Based on this, we estimate that in Case~1, after PNAS prohibited editors from handling submissions by any collaborators from the past 24 months, the number of papers with editor-author collaboration within that time span decreased from 10.32\% to 8.49\% ($p = 0.029$). Moreover, the percentage of papers with such editor-author collaboration continued to decrease by about 0.5\% per year during the five-year period that followed the policy change ($p < 0.001$).
{Extrapolating the pre-policy trend, we estimate that 13.3\% of papers would have involved an editor–author collaboration within the preceding 24 months, had the new policy not been adopted. This is nearly double the regression-estimated value based on actual observations, which is just 6.9\% (Supplementary Figure~7). Put differently, we estimate that the policy led to a 6.4 percentage point reduction in such collaborations four years after its implementation.}
In the other two cases, we found no evidence that the policies had any effect on the percentage of papers with recent editor-author collaborations; see Supplementary Table~8 for regression estimates. {These findings suggest that, despite policies forbidding editor-author associations, such papers continue to be published without any apparent repercussions. For instance, out of all papers that we identified as having an editor-author association in PNAS and PLOS, only three were retracted citing these associations as reasons~\cite{retraction_watch}. The lack of corrective procedures associated with the vast majority of such papers could explain the ineffectiveness of current COI policies.}

As is the case with any observational study, an RDiT design, such as ours, comes with some intrinsic limitations that should be carefully considered. Firstly, there is often a need to expand the window (i.e., the period before and after the treatment) in order to obtain sufficient statistical power~\cite{hausman2018regression}. However, by expanding the window, it becomes harder to attribute any observed change to the treatment in question. One way to alleviate this issue is to perform a sensitivity analysis while varying the window size. Accordingly, we adopt alternative specifications where the bandwidth around the cutoff date is varied; this yields similar results (Supplementary Table~9). {Here, we do not opt for thresholds smaller than 36 months to ensure sufficient statistical power. Consequently, our results rely on observations that may be considered relatively far from the threshold, even with the minimum bandwidth (i.e., 36 months).}

Secondly, the observed change (or lack thereof) in the outcome around the treatment time could be attributed to other events that coincide with the treatment. To rule out this possibility, we use a negative control group for each of the three policy changes. These groups consist of papers whose author(s) collaborated with the handling editor, but the collaboration fell outside the range specified by the policy in question. Fig.~\ref{policy_change}d to \ref{policy_change}f depicts the control groups of Cases~1 to 3, respectively. In Case~1, for example, the treatment group (Fig.~\ref{policy_change}a) involves collaborations within the past 24 months, while the control group (Fig.~\ref{policy_change}d) involves those within the past 25 to 48 months. Hence, the treatment and control groups are influenced by the same exogenous factors (if any) apart from the policy change, while the latter applies only to the treatment group. Now if the observed pattern around the cutoff date in the treatment group is attributed (at least partially) to the policy change, we would expect to see a different pattern in the control group. However, both groups differ only in Case~1 (see {Supplementary Table~8}), suggesting that the decrease in {the percentage of papers with editor-author collaboration} in Case~1 is likely due to policy change, and that the policy likely did not have any effect on the {percentage of papers with editor-author collaboration} in Case~2 and 3. 

\vspace{0.5cm}
\subsection*{The suitability-integrity tradeoff of managing editor-author association}
There could be many reasons why recusal policy is not working as intended. Here, we consider one of the biggest reason---suitability. Note that editors who are selected to handle a paper are often those whose field resembles that of the paper~\cite{PNASpolicy,resnik2016ensuring,PLOSPeerReviewPolicy,HindawiPeerReviewPolicy}. However, those are arguably the editors who are more likely to have a professional relationship with the authors. Based on this observation, some have argued that prohibiting all {editor-author associations} could compromise peer review quality, since the most suitable editors (in terms of expertise) may be prohibited from handling the paper in question~\cite{resnik2018conflict,gottlieb2017should}. However, this argument has not been put to the test to date.
To this end, we use a graph embedding model to encode the fields of research of papers and scientists in a high dimensional space; see Methods for details on how the embeddings are calculated. This way, for any given paper $p$ handled by editor $e$, we are able to determine whether there are editors on the same editorial board as $e$ who could have handled $p$ given their expertise.

To begin with, for any given paper $p$ handled by editor $e$ and published in journal $j$, let us examine the difference in expertise between $e$ and a random editor $e'$ serving on the same editorial board. Naturally, we would expect the expertise similarity between $p$ and $e$ to be greater than that between $p$ and $e'$, since the handling editor tends to be the most relevant to the paper in terms of expertise. Similarly, we would expect the similarity between $p$ and $e'$ to be greater than that between $p$ and a random editor in our dataset, who may belong to an entirely different discipline. Fig.~\ref{expertise}a shows that this is indeed the case, suggesting that our embeddings are able to capture the expertise of editors and papers.

Next, we examine the relationship between {editor-author associations} and expertise. Fig.~\ref{expertise}b shows that the {probability of having editor-author associations} positively correlates with expertise similarity, i.e., the more similar the expertise of a paper is to that of its handling editor, the more likely it is for the handling editor to have {an association with the authors}. Importantly, when the expertise similarity between a paper and its handling editor is above 0.96 (the 95-th percentile value), we estimate that 17.12\% of editors have {an association with the authors} (95\%-CI is [16.67\%, 17.56\%]). This suggests that, if journals were to enforce COI policies, then the editor who ends up handling the paper may not be the most relevant on the editorial board in terms of expertise. To determine whether this is the case, instead of comparing the handling editor $e$ to a random member of the editorial board $e'$, we need to compare $e$ to the most relevant editorial board member (in terms of expertise) who does not have {any association with the authors}, denoted as $e^*$.

As can be seen in Fig.~\ref{expertise}c, for the majority of papers (over 70\%), the expertise similarity between $e^*$ and $p$ is higher than that between $e$ and $p$. This analysis suggests that, in 70\% of cases, the paper in question could have been handled by an alternative member of the editorial board who is not only more suitable (in terms of expertise), but also has no associations with any authors. Viewed from a different perspective, these findings suggest that 30\% of cases can only be resolved by compromising the suitability of the handling editor. {Recall that the team size of prior collaborations plays a role in the governance of potential COIs (see Supplementary Figure~5). Therefore, we further examine how this percentage would change when we account for the size of prior collaborations. As shown in Supplementary Figure~8, if policies require the recusal of the handling editor if there existed a prior collaboration involving no more than 20 co-authors, 25.5\% of papers would still be handled by less suitable editors. This percentage becomes 21.11\% if the threshold of prior team size is chosen to be 10. This suggests that a sizable portion of papers would still be handled by less suitable editors, even if recusal policies adopt more conservative definitions of COI based on the team size of prior collaboration.}

Intrigued by the above findings, we further examine the trade-off between avoiding {editor-author associations} and assigning the most suitable editor. To this end, we adopt the same RDiT design used earlier in our examination of policy impact, and use it to estimate how PNAS's policy change affected handling editors' suitability. Recall that this policy change reduced the {percentage of papers with editor-author associations} in PNAS soon after its introduction (Fig.~\ref{policy_change}a). Despite such reduction, however, we find no evidence that the policy change affected the average expertise similarity between papers and their handling editors (Fig.~\ref{expertise}d and Supplementary Table~10).

{
\subsection*{Exploring public disclosure as an alternative approach to govern editors' COI}
It has been argued that the existence, or even perception, of COI could decrease the public's trust in science~\cite{WAME2009COI,ICMJE,friedman2002impact}, yet it is unclear how to best govern potential COIs stemming from editor-author associations. In particular, our previous analyses suggest that current COI policies have limited effect in regulating editor-author associations, but also suggest that a blanket restriction to prohibit all editor-author associations may compromise the suitability of a sizable portion of handling editors.
These findings call for alternative approaches to manage such potential COIs, as opposed to the more common practice of recusal currently employed by most publishers. With this in mind, we consider adding to the paper a public disclosure that explicitly acknowledges the potential COI along with any measures taken to ensure impartiality, e.g., by appointing a secondary editor to oversee the handling the process. Although public disclosure has already been deployed widely to govern author-related COIs~\cite{malivcki2021systematic}, it has not been adopted for editor-related COIs to date. While some publishers require editors to declare their competing interests, such declarations are never made public. Similarly, while some journals appoint a secondary editor (often editor-in-chief) to oversee editorial decision when a manuscript is submitted from an editorial board member (for example, \cite{jpubhelthpolicy}), this policy is not applied to govern editor-author associations. 


To assess the viability of the proposed measure as an alternative governance of editors' COIs, we designed a preregistered survey experiment to measure the effect of the aforementioned disclosure on the general public's trust in science. In this survey, respondents are asked to first read descriptions of three hypothetical scientific articles, each consisting of a summary of the finding, followed by a short bio of the author. Then, they are asked to rate their level of trust in each finding and each author on a 7-point likert scale (Fig.~\ref{survey}a). The three articles are presented in random order. Importantly, when presenting the middle one, participants are informed that the study involves a COI due to an editor-author association. Depending on how the COI information is worded, respondents are randomly assigned to two conditions: (i) No-disclosure condition: Here, respondents are informed that the COI is not publicly disclosed; (ii) Disclosure condition: Here, respondents are presented with a statement openly disclosing the editor-author association along with measures taken to ensure impartiality; see Methods for more details regarding the survey design, and see Supplementary Note~3 for survey vignettes corresponding to either condition.

Arguably, one would expect that reading about a COI would reduce the reader's trust in the findings. It is also plausible that reading the disclaimer would reduce the negative effect of the COI. However, what is not obvious is whether the disclaimer would entirely negate this effect. The primary purpose of our randomized controlled trial is to address this question. To this end, we first test whether knowledge about the COI decreases the respondents' trust in the finding and/or the author. In particular, we compare trust level reported towards the first article and the second article by respondents in the no-disclosure condition. As can be seen in Fig.~\ref{survey}b, knowing about the COI significantly reduced the respondent's trust in the scientific finding ($t_{454} = 4.313$, $P = 1.974e-05$, Cohen's $d = 0.161$) but had no significant effect on the trust level towards the author of that finding ($t_{454} = -0.832$, $P = 0.406$, Cohen's $d = 0.035$). Next, we investigate the extent to which public disclosure of COIs negates the aforementioned effect. To this end, we compare the difference in differences between trust level reported towards the second article and that of the first article among respondents of both conditions. As can be seen in Fig.~\ref{survey}c, we find that the public disclosure of editor's COI yields a significantly higher level of trust towards the finding ($t_{908} = -3.316$, $P = 9.486e-4$, Cohen's $d = 0.220$) but not the author ($t_{908} = -2.159$, $P = 0.031$, Cohen's $d = 0.143$).

Now that we have shown that COI has a negative effect on trust, and disclosure has a positive effect, what remains to be answered is whether these effects cancel each other out. To address this question, we conduct an additional, non-preregistered analysis that focuses on the disclosure condition, comparing the trust levels between the first article (serving as the baseline) and the second article (which included the public disclosure). As can be seen in Fig.~\ref{survey}d, trust in the scientific finding with a disclosed COI is not significantly different to the baseline article ($t_{454} = -0.834$, $P = 0.404$, Cohen's $d = 0.038$), while trust in the author involved in the disclosed COI is significantly higher ($t_{454} = -3.638$, $P = 3.060e-4$, Cohen's $d = 0.160$). These results provide evidence that public disclosure of editors' COIs could enhance trust in authors without compromising trust in scientific findings, suggesting that our proposed approach of governing editor-author associations comes with substantial benefits and no measurable drawbacks.

We already established that trust in a scientific finding decreases when respondents become aware of a COI due to an editor-author association. But does this effect extend beyond the scientific finding in question? In other words, if a particular paper involved such a COI, does it influence the reputation of other papers even if those papers did not involve any COI? To explore this possibility, we compared the respondents' trust towards the first and third articles. Notice that both articles do not involve any COI, with the only difference being that respondents went over the first article without reading about any COI, but went over the third one having read about a COI associated with the second article. Our results reveal that the levels of trust towards the third article are not statistically significantly different from those of the first article in all analyses (Fig.~\ref{survey}b through \ref{survey}g). This suggests that both the effect of COIs and the effect of their disclosure are localized to the articles directly associated with the COI. 
}

\section*{Discussions}


In this study, we demonstrated that editors often handle papers (co-)authored by their recent collaborators or by their colleagues who share the same affiliation. 
Additionally, we demonstrated that these papers are accepted faster only when the prior editor-author collaboration happens in relatively small teams, raising the possibility of favoritism in those cases.
{Moreover, we showed that papers with editor-author associations are not just concentrated in special issues or low-quality journals. In fact, the majority of journals with the highest editor-author associations rates are actually Q1 journals, and editors with higher ranked affiliations are more likely to be associated with the authors.}
Leveraging three cases of policy change as quasi-experiments, we found that COI policies may have an effect on regulating papers with editor-author associations, even when they are not enforced in practice. However, enforcing such policies will compromise the suitability of the handling editors of some submissions. We estimated that while 70\% of papers with editor-author associations can be handled by alternative suitable editors, enforcing such policies might result in 30\% of such submissions being handled by less suitable editors, revealing a suitability-integrity trade-off when enforcing COI policies.

Our study, however, is not without limitations. One limitation stems from the fact that we focus on the publishers that openly share data on the handling editors of all papers, rather than analyzing a random sample of journals. Hence, each analyzed publisher has its own distinguished characteristics that may prevent the generalization of our findings. Specifically, Open Access publishers such as Frontiers, Hindawi, and MDPI attract scrutiny due to large volumes of Special Issues published therein~\cite{nicholas2023never}, while PNAS has ``inside tracks'' that facilitate the publication of manuscripts submitted by members of the U.S.~National Academy of Sciences~\cite{kean2009pnas}. We accounted for these specific characteristics by comparing special issues to normal ones in Frontiers, Hindawi, and MDPI, and by excluding inside tracks from the analysis of PNAS. Having said that, a random sample of journals could be more informative. Unfortunately, this is not possible due to the lack of publicly available data specifying the handling editor of each paper. There are a few publishers{, such as Elsevier,} who selectively publicize such information in a small fraction of their journals, but these journals are not necessarily representative of all the journals handled by that publisher. {Furthermore, upon reaching} out to the top 30 publishers to request such data for analysis~\cite{nishikawa2022100}, {the few that responded denied our request}. This highlights the need for greater transparency in sharing data related to editorial processes.

The second limitation is due to the fact that we infer editor-author associations using publication records only. Therefore, future studies should consider other types of editor-author association that are explicitly mentioned in the publishers' policies. Examples include situations where an author of the manuscript under consideration has previously written a grant proposal with the editor, or where an author has {previously} been a member of the editor's research lab, either as a PhD student or as a postdoc. {While some} traces of such relationships may already be captured by co-authorships in our dataset, additional datasets of grant proposals and mentor-mentee relationships would provide more insights into this form of editor-author association.

{The third limitation concerns the survey design. 
Respondents were presented with hypothetical scientific articles from the Social and Behavioral Sciences, which were reviewed by researchers to ensure plausibility. While these topics were chosen for their public relevance, it remains unclear whether our findings would generalize to other scientific fields. Additionally, our analysis focuses specifically on how COIs influence trust among members of the general public. This choice was deliberate, since concerns about the public’s trust of science are often cited as a key rationale for COI governance~\cite{ICMJE,WAME2009COI,friedman2002impact}. Nevertheless, it is interesting for future research to extend this work by examining whether similar effects are observed among scientists and domain experts. Finally, because respondents rated their level of trust immediately after reading each article, the study captures only short-term effects. It remains an open question whether the effects of COI disclosures on trust persist over time.}

The fourth and final limitation is that, due to the quantitative nature of our study, we do not take into consideration the nuanced situations in which each manuscript is handled. As we have documented in Supplementary Note~2, not all editors perform the same roles in the peer review process; some publishers allow the handling editors to decide the reviewers and overrule their recommendations, while others do not. Therefore, having the same type of editor-author association may carry different implications across publishers. {This heterogeneity complicates the interpretation of our findings. For instance, while our results suggest that 30\% of papers would need to be reassigned to less suitable editors in order to eliminate all potential COIs (Figure\ref{expertise}c), it remains unclear whether this tradeoff is acceptable or consequential in practice. More research is needed to evaluate how replacing editors with potential COIs affects peer review quality, publication outcomes, and long-term trust in the editorial process.}


There is a wide spectrum of opinions regarding how non-financial COIs should be perceived. On one end of the spectrum, when commenting on the first ever retraction from PNAS involving an editor's COI, the editor-in-chief stated that COI alone would have been sufficient to prompt a retraction, regardless of the correctness of the scientific findings~\cite{retraction2022PNAS}. On the other end of the spectrum, a perspective piece published by PLOS argued that having a non-financial COI is not a COI at all~\cite{bero2016having}. Our findings indicate that both opinions are limited by demonstrating that the reality of COI is far more complicated---collaborations that take place in different contexts may carry different implications, and policy-makers need to consider the integrity-suitability tradeoff that was quantified for the first time in this study.
{In light of these complexities, our study identifies key limitations in current COI policies and raises two critical questions for better governance of COI.}

The first question is {whether it is possible to clearly define} the types of editor-author associations that constitue COI. After all, numerous factors affect how COIs are perceived under different circumstances. For instance, an editor-author collaboration that took place a decade ago may not raise the same concern as another that took place only a year ago. Similarly, an editor-author collaboration involving 100 co-authors may not be perceived in the same way as one involving a smaller, close-knit team. While most current policies recognize the former factor, none of them recognize the latter.

Our analysis of relative acceptance delay suggests that team size matters. More specifically, our findings raise the possibility of favoritism, but only when the editor-author collaboration involves a team of fewer than 20 co-authors (Fig.~\ref{rad}d and Supplementary Figure~5). 
{This may be because large teams are often organized as research consortia that lack direct interaction among all contributing authors, especially when the collaboration spans multiple institutions and countries~\cite{coles2022build,katz1997research}. In such teams, co-authorship can be granted to people who merely contribute to the shared project infrastructure~\cite{thelwall2020large,authorship2008}, unlike smaller teams where co-authorship indicates substantial contributions to the paper~\cite{maddi2024beyond}. Therefore, although co-authorship has traditionally been treated as synonymous with research collaboration, current evidence highlights the need to re-evaluate this equivalence in the age of ``big-team science''~\cite{smith1958trend,melin1996studying,wagner2005network,lerner2023micro}. 
While it is empirically challenging to distinguish between co-authorship and de facto research collaboration, our findings indicate that the size of prior teams should be taken into account when governing conflicts of interest.}

Other than team size, many other factors could, in principle, also be considered. For example, what was the role of the authors who previously collaborated with the editor? Were they the first or last author, or simply a middle author? Additionally, did the editor-author collaboration produce a conventional research paper, or a white paper published in a scientific journal? Moreover, personal associations take many forms; not all COIs arise from being a colleague, collaborator, mentor, and other roles defined by the current policies. Apart from these tangible influences, there are more subtle and elusive ones, such as intellectual COIs~\cite{luborsky1999researcher}. These examples demonstrate the intricacies involved when defining COIs, and suggest that producing a comprehensive list of COI sources---a prerequisite for enforceable recusal policies---may not be practical.

This brings us to the second question: What is the goal of a COI policy? Drawing on our findings, we argue that any COI policy should balance three primary concerns: fairness, suitability, and public trust. First, fairness is critical because editors, being human, are susceptible to biases. 
As for the second concern, suitability must be balanced against fairness. If one were to eliminate all editor-author associations, the suitability of editors might be compromised, since those who are most suitable to handle a manuscript are arguably more likely to have some form of associations with its authors~\cite{resnik2018conflict}. As our analysis has demonstrated, this is indeed a legitimate concern, as it suggests that 30\% of papers with editor-author association would have otherwise been handled by a less-suitable editor.
One possible way around this tradeoff could be to invite guest editors when no other editorial board member is suitable to handle the paper in question, according to the editorial policy of many publishers~\cite{PNASpolicy,PLOSPeerReviewPolicy,HindawiPeerReviewPolicy,MDPIPeerReviewPolicy}, but this would increase the burdens of editorial board members, and risk further slowing down the peer review process. {Employing professional editors, i.e., academics who cease academic publishing and become full-time editors (such as those employed by \textit{Cell}, \textit{Nature}, \textit{Science}, and \textit{Nature Communications}), would also in principle reduce editorial COIs, but it comes with additional costs that not all publishers and academic societies can bear.}
As for the third concern, public trust is at stake when governing editors' COI~\cite{WAME2009COI,ICMJE,friedman2002impact}. Indeed, our survey experiment demonstrates that informing a reader of the editor-author association can hurt readers' trust in the scientific findings.
The broader implication of our experiment is that public trust can be restored when measures to ensure impartiality are taken and disclosed transparently.


Taken together, our results challenge whether recusal is a suitable approach for governing editor's COI, and instead advocate for public disclosure as an alternative. {However, relying on voluntary disclosure alone to detect such cases is unreliable, especially since editors may lack the incentive to fully disclose all potential COIs~\cite{taheri2021discrepancies}.
Consequently, publishers should better integrate their manuscript tracking system with bibliometric databases. As demonstrated in our study, recent collaborations and shared affiliations---the two most widely recognized sources of editorial COI---can be automatically identified using such databases. This way, publishers could mandate disclosure when editor-author ties are flagged in the system. Furthermore, there should be repercussions of non-compliance. If undisclosed COIs are exposed, corrective actions could include retraction regardless of scientific merit (as endorsed by the editor-in-chief of PNAS~\cite{retraction2022PNAS}), post publication peer review, or issuing a notice of concern (e.g., see~\cite{maeyama2014palindromic}). Our survey suggests that readers consider such notices when assessing their trust of the article in question. Moreover, the effect is localized and does not extend to other articles without similar disclosures.

When defining what should be disclosed, inconsistency across current editorial policies (as demonstrated in Table~\ref{tab:policy} and \ref{tab:publishers}) presents a challenge. One possible way to circumvent this is to adopt the union of all items listed across those policies to ensure comprehensive coverage. Furthermore, our analysis suggests that, in addition to the recency between prior collaboration and manuscript submission, the nature of prior collaboration should be disclosed, since our findings suggest that teams larger than certain discipline-specific thresholds may not necessarily introduce COIs.}

Beyond safeguarding trust in science, transparent disclosure offers a multitude of benefits. First, preparation of the disclosure requires editors to consciously examine their relationship with the authors and the manuscript in question, potentially leading to fairer decisions. Second, disclosures allow readers to contextualize scientific findings within the broader social dynamics shaping their production. {Last but not least, transparent disclosures will generate a wealth of data which enable quantitative studies of the epistemic impact of COIs on science.} Such analyses could identify the ways in which social relationships advance or hinder scientific progress, providing additional evidence to inform how policies should be designed to best manage such relationships.

\section*{Methods}
\subsection*{Ethics}
The survey experiment was approved by the New York University Abu Dhabi Institutional Review Board (IRB) with study number HRPP-2024-141. The rest of studies are observational using publicly available data, so no approval was requested from the IRB. A formal exemption from the New York University Abu Dhabi IRB (HRPP-2024-55), however, was later obtained during the peer review process. The observational studies were exempted due to ``No Human Subjects Research.''

Here, we emphasize that COI is a complicated phenomenon, shaped by many external factors such as peer pressure, editorial duties, research culture, or the lack of alternative editors who have the necessary background to handle the paper in question. Additionally, scientific research increasingly {takes} place in large teams, involving independent research groups in different countries{, which challenges the notion that prior collaboration constitutes a source of COI}. Moreover, other factors, such as the rapid expansion of research institutes and the tendency of scientists to hold multiple, sometimes non-contractual appointments, further complicates the definition of COI. As such, accusatory language should be avoided when interpreting our findings.

In order to preserve the anonymity of editors, we do not release the dataset that is being analyzed in this study. Those who are interested in replicating our analysis can download all datasets required for this study from the Internet, following the procedures described in the Methods section of Liu et al.~\cite{liu2023non}. Due to the sensitive nature of the research question, those who reproduce our research should adhere to the highest standard to avoid revealing the identities of the involved editors. Naturally, any attempt at reproducing or extending our analysis should be subjected to the approval of their respective institutional review boards.

\subsection*{COI policies}
This study focuses on six publishers: Public Library of Science (PLOS), Frontiers Media S.A. (Frontiers), the Multidisciplinary Digital Publishing Institute (MDPI), Hindawi Publishing Corporation (Hindawi), the Institute of Electrical and Electronics Engineers (IEEE), and the Proceedings of the National Academy of Sciences (PNAS). Although we refer to them as ``publishers'' in an encompassing way, they are in fact four publishers, one academic society (IEEE), and one multidisciplinary journal (PNAS).

To collect the COI policy of each publisher, we visited the home page of each publisher on the Internet, and looked for hyperlinks related to ``publication ethics'' or ``journal policies'' on their home pages. For example, the home page of Frontiers contains a link to the ``Policies and publication ethics'' page which contains a section related to COI. Similarly, the home page of PLOS contains a drop-down menu that links to ``competing interest'' under the ``Policies'' section, the home page of MDPI contains a link to ``Research and Publication Ethics'' page under the ``Information'' dropdown menu, the home page of IEEE contains a link to ``Organizational Ethics'', and the home page of PNAS contains a link to ``Editorial and Journal Policies'' page in the ``Authors'' drop-down menu. As for Hindawi, since we could not find ethics- or policies- related links on its home page, we first visited the ``Editor'' page of Hindawi, and then identified the ``Editor Code of Conduct'' page that contains information related to COI.

In this study, we also analyzed historical versions of COI policies from PNAS and PLOS; see the Supplementary Note~1 for details.

{
\subsection*{Interpretation of policies}
As can be seen in Table 1, different COI policies adopt different modal verbs (such as ``may'', ``should'', or ``must'') when it comes to expressing the recommendation, permission, or obligation regarding whether editors should recuse themselves when a COI is present. Depending on the modal verbs, we classify the publishers into three tiers. The first consists of PLOS and Frontiers, whereby both do not require nor recommend editors to recuse themselves when they have potential COIs, and simply offer recusal as an option. The second tier consists of IEEE, MDPI, and Hindawi, which do not recommend editors handling papers with a COI. These publishers state that editors ``should decline'' to edit or ``should not'' edit such papers. The third tier, represented by PNAS, employs the strictest language, stating that editors ``may not'' handle papers with COI. The phrase ``may not'' typically expresses lack of permission or absolute prohibition~\cite{maydefinition}, making it more stringent than ``should not''~\cite{shoulddefinition}. Notice, however, that PNAS does not consider editor-author sharing the same affiliation as a source of COI.
}

\subsection*{Dataset}
The name and affiliation of editors, as well as the papers that they handle, are obtained in the same way as in Liu et al.~\cite{liu2023non}. These editors come from four publishers (PLOS, Frontiers, Hindawi, and MDPI), one academic society (IEEE), and one multidisciplinary journal (PNAS). Following the methods described therein, we did not analyze all papers published in PNAS but only analyzed papers submitted through the direct submission track, which makes up the vast majority of PNAS papers, and excluded from our analysis all communicated papers (a submission track that was discontinued in 2010) as well as all contributed papers (a submission track that only members of the National Academy of Sciences can use).

Next, we disambiguate editors by identifying their publication records. To do so, we need to collect the names and the affiliation of the handling editor of each paper.
In all journals analyzed, the handling editor's name is published along with each paper. As for the editors' affiliation, this information is published along with each paper in PLOS, Frontiers, PNAS, and IEEE, which means that no extra processing is needed for those publishers. {Since the affiliation of an editor is reported along with each paper that one handles, any changes in an editor's affiliation is reflected in our dataset.} However, Hindawi and MDPI only specify {the affiliations of} the current members of each editorial board. As such, we downloaded past versions of the editorial board web pages using the Wayback Machine~\cite{wayback} and then recorded the affiliations of the editors listed in each version. {Wayback Machine allows us to retrieve historical affiliation data of editors that tracks the changes in affiliation that editors may have had.}

Having collected the name of each editor, as well as the year in which an editor is affiliated with a certain affiliation, we disambiguate the editors using Microsoft Academic Graph (MAG), a bibliographical dataset of over 200 million published papers and preprints~\cite{sinha2015overview, wang2019review}. {We use a local snapshot of MAG that we downloaded in December 2021, before public access to MAG stopped. This local snapshot of MAG covers all years spanned by our editor dataset.} Due to the shear volume of information recorded by MAG, it may contain erroneous entries of publications. However, it is a trusted source of bibliography due to its conservative approach to the name disambiguation task, which favors underconflation to overconflation~\cite{wang2020microsoft}. In other words, for any two authors that share the exact same name, they are recorded as two different entities in the dataset unless the probability of them being the same person is extremely high. Consequently, the MAG team states that they ``have confidence that when Microsoft Academic attributes a set of papers to an author, they were actually written by that person''~\cite{wang2020microsoft,wang2019review}. Following Liu et al.~\cite{liu2023non}, we adopt the following rule when disambiguating editors using MAG: for each editor $e$ affiliated with university $u$ in year $t$, the editor $e$ is identified to be the same person as scientist $s$ in MAG if and only if $s$ happens to be the only scientist in MAG with the exact same name as $e$, and is affiliated with university $u$ in year $t$. Also adopting a conservative approach, we omit papers whose editor cannot be found in MAG or cannot be uniquely matched with a scientist in MAG.

\subsection*{Defining recent collaborations}
For a given paper $p$ authored by a set of authors $\mathcal{A} = \{a_1, a_2, \dots, a_n\}$ and edited by a handling editor $e$, if there exists $a_i \in \mathcal{A}$ such that $a_i$ and $e$ have co-authored a paper $q$ prior to the submission of $p$, then $\Delta$ of $p$ is defined to be the difference in the number of months between the submission of $p$ and the publication of $q$. Formally, $\Delta = t_p - t_q$, where $t_p$ is the date on which $p$ is submitted, and $t_q$ is the date on which $q$ is published. An author of $p$ has a recent collaboration with $e$ if and only if $\Delta \leq t$ where $t$ is any threshold identified by the publisher of $p$.

{Empirically, publishers differ in their definition of $t$, ranging between 24 months (or 2 years) and 60 months (or 5 years), while some publishers do not explicitly define $t$ at all. Given that no consensus exists over what period constitutes ``recent'', we pick a middle value of $t = 48m$ in our analysis. Additionally, we report the percentage of papers with recent editor-author collaborations in Supplementary Figure~1 if alternative values of $t$ are adopted (i.e., $t = 24m$, $t = 36m$, and $t = 60m$).}

{
\subsection*{Percentage of papers with editor-author associations}
This study considers two types of editor-author associations: editor-author association due to recent collaboration and editor-author association due to sharing the same affiliation. Therefore, if a paper has both types of editor-author associations, it contributes to both Fig.~\ref{coi_rate}a and Fig.~\ref{coi_rate}b.}

Following Liu et al.~\cite{liu2023non}, we excluded from our analysis papers submitted through the ``inside tracks'' of PNAS. More specifically, we excluded contributed papers (i.e., those submitted by members of the National Academy of Sciences along with peer review reports from referees designated by the member) as well as communicated papers (i.e., those submitted in private to an Academy member who first secures reviewer feedback before officially submitting the paper to PNAS)~\cite{kean2009pnas}.

We complement Liu et al.'s work~\cite{liu2023non} by additionally gathering the information of whether each paper belongs to a special issue in Hindawi and MDPI, or a ``research topic'' (akin to a special issue) in Frontiers. Each journal in Hindawi curates its own collection of special issues. Similarly, each MDPI journal also curates its own list of special issues, while other special issues may belong to a ``section''. As for Frontiers, all research topics published therein are readily available in a single webpage.

Finally, the university ranking data is sourced from the Academic Ranking of World Universities, or commonly known as the ``Shanghai Ranking''~\cite{shanghairanking}. It specifies the world's top 1000 research universities, and ranks them primarily by their research outputs, e.g., awards, publications, citations, etc.~\cite{shanghairankingmethod}. Therefore, all other research institutions not present on this list are coded as ``1000+'' in ranking in our analysis. One limitation of this ranking lies in the absence of departmental-level information.

{
\subsection*{Calculating relative acceptance delay}

In Fig.~\ref{rad}, we compare the papers with and without recent editor-author collaboration in terms of their acceptance delay, i.e., the number of days spent between submission and acceptance, while accounting for the temporal and cross-sectional variation in acceptance delay. To do so, for each paper $p$, published in journal $j$ and year $y$, we compute the relative acceptance delay (RAD), measured as the difference in acceptance delay between $p$ and the average paper published in $j$ in year $y$.

\subsection*{Regression discontinuity in time}
In Fig.~\ref{policy_change}, each data point represent the monthly percentage of papers with recent editor-author collaborations in the publisher that was studied, which was calculated by dividing the number of published papers with recent editor-author collaborations that was submitted in a certain month by the total number of published papers that was submitted in that month. We used the following regression specifications to estimate the temporal trend as well as the effect of COI policies: percentage = Month + Policy + Month $\times$ Policy, where ``Month'' represents the number of months before the cutoff date (if it is a negative value) or after the cutoff date (if it is a positive value), and ``Policy'' is a binary variable, which is set to be ``true'' if the current month is after the cutoff date of the policy change and set to be ``false'' otherwise. The interaction term captures how the temporal trend changed after the policy update.}

\subsection*{Approximating author expertise}
{Although it might be difficult to find editors whose expertise aligns with manuscripts in certain multidisciplinary journals such as PNAS and PLOS One, in general,} editors are selected to handle manuscripts similar to their own areas of expertise~\cite{PNASpolicy,PLOSPeerReviewPolicy,HindawiPeerReviewPolicy}. {This suggests that the expertise of the handling editor of a manuscript is expected to be close to that of its authors, at least relative to most other editors who did not handle this paper.} Arguably, the proof for having expertise in a research field comes from a strong publication record in that field~\cite{radun2023nonfinancial}. One way to encode the similarities between research fields is to use network embeddings, which have been shown to outperform their text embedding counterparts for this purpose~\cite{constantino2023representing}.

Against this background, we estimate the expertise of editors and authors using MAG's citation network.
We then apply a node2vec algorithm to calculate network embeddings using a scalable implementation of the algorithm~\cite{cappelletti2023grape}.
The expertise representation of a scientist is then calculated as the average embedding of all papers that they have (co-)authored. The similarity between an editor and a paper is calculated as the cosine similarity of their vector representations in the expertise space.

{
\subsection*{Survey}
The survey study was preregistered on the AsPredicted platform on October 4, 2024 under registration number \#192642. The preregistration report is made publicly available at \url{https://aspredicted.org/6yr4-db7s.pdf}. The survey is designed using Qualtrics, and the participants are recruited on Amazon Mechanical Turk (MTurk). At the end of the survey, we collected demographics information of the participants, which is reported in the Supplementary Materials. Each survey respondent was provided with a randomly generated token after completing the survey, which they submit on Aamazon MTurk so that we can use this token to link survey responses with MTurk records. We included comprehension check questions throughout the survey to ensure that participants have actually read the texts displayed to them; those who fail the comprehension tests are automatically discontinued from the survey and are excluded from our analysis. We test eight hypotheses in total using the survey experiment (see pre-registration report for details). To correct for mutiple hypotheses testing, we use the Bonferroni correction and set the corrected p-value at 0.00625 to ensure that the overall probability of a Type I error (false positive) rate of 5\%. A pilot study of 300 participants was conducted in November. Based on results in the pilot study, we estimated that the effect size to be small, around 0.2 (Cohen's $d$). Setting the desirable power at 0.8, we perform the power analysis that leads to the number of participants to be 455 per condition (910 participants in total). Due to the unpredictability of the passing rate of comprehension checks, we initially surveyed more respondents than required, but we only analyze the first 455 respondents per condition to ensure the per-determined Type I and Type II error rate.
}

\section*{Acknowledgments}
F.L. is supported by the New York University Abu Dhabi Global PhD Student Fellowship. The support and resources from the High Performance Computing Center at New York University Abu Dhabi are gratefully acknowledged. We thank Dashun Wang, Yong-Yeol (YY) Ahn, and anonymous reviewers for their valuable suggestions. 

\section*{Data and code availability}
The anonymized data and all custom code used to analyze those data is provided \href{https://github.com/Michael98Liu/editor-author-association}{here}.

\newpage
\begin{table}[]
\resizebox{\linewidth}{!}{%
\begin{tabular}{p{2cm}p{5cm}|p{4.5cm}p{4.5cm}p{4.5cm}p{4.5cm}}
\hline
\hline
 &  & \textbf{COPE~\cite{COPE_COI}} & \textbf{CSE~\cite{CSE}} & \textbf{ICMJE~\cite{ICMJE}} & \textbf{WAME~\cite{WAME2009COI}} \\
 \hline
 \hline
\multicolumn{1}{c}{} & Stringency of recusal policy & Medium & Medium & Medium & Medium \\
\multicolumn{1}{c}{} & Ask for disclosure & Yes & Yes & Yes & Yes  \\
\multicolumn{1}{c}{\multirow{-2}{*}{\shortstack[l]{General policy\\regarding editor-\\author associations}}} & Other non-financial COIs & Mentor/mentee, joint grant holder & Competitors & N.A. & Family members, friends, enemies, competitors \\
\hline
\multicolumn{1}{c}{} & Mentioned in the policy & Yes & Yes & No & No \\
\multicolumn{1}{c}{} & Only recent collaboration & Yes & Yes & N.A. & N.A. \\
\multicolumn{1}{c}{} & Clear definintion of ``recent'' & 3 years & No & N.A. & N.A. \\
\multicolumn{1}{c}{\multirow{-4}{*}{\shortstack[l]{Editor-author\\associations due to\\recent collaboration}}} & Additional contraints & close collaborator & No & N.A. & N.A. \\
\hline
 & Mentioned in the policy & Yes & Yes & No & Yes \\
 & Same affiliation current & Yes & No & N.A. & No  \\
 & Same affiliation in the past & No & No & N.A. & No \\
\multirow{-4}{*}{\shortstack[l]{Editor-author\\ associations due to\\same affiliation}} & Additional conditions & No & Same department only & N.A. & Same department or ``small'' institutions 
\end{tabular}%
}
\caption{Overview of COI policies adopted by academic organizations. Here, policy ``stringency'' refers to the degree of freedom that editors have when deciding their action, when they face a certain type of editor-author association that is specified in the policy; more details regarding how stringency level is coded can be found in Methods. Note that COPE does not formally adopt guidelines concerning editors' COI. However, according to the COPE Council, COPE's ethical guidelines for peer reviewers~\cite{COPE_COI} are also applicable to handling editors~\cite{COPE_editor_COI}.}
\label{tab:policy}
\end{table}

\begin{table}[]
\resizebox{\linewidth}{!}{%
\begin{tabular}{p{2cm}p{5cm}|p{3cm}p{3.5cm}p{1.7cm}p{3cm}p{2.8cm}p{4cm}}
\hline
\hline
 &  &  \textbf{Frontiers~\cite{FrontiersPolicy}} & \textbf{Hindawi~\cite{HindawiPolicy}} & \textbf{IEEE~\cite{IEEEpolicy}} & \textbf{MDPI~\cite{MDPIpolicy}} & \textbf{PLOS~\cite{PLOSpolicy}} & \textbf{PNAS~\cite{PNASpolicy}} \\
 \hline
 \hline
\multicolumn{1}{c}{} & Stringency of recusal policy & Low & Medium & Medium & Medium & Low & High \\
\multicolumn{1}{c}{} & Ask for disclosure & Yes & Yes & Yes & Yes & Yes & Yes \\
\multicolumn{1}{c}{\multirow{-1}{*}{\shortstack[l]{General policy\\regarding editor-\\author associations}}} & Other non-financial COIs &  Family members & Have a close personal connection to any author; feel unable to be objective & N.A. & Personal friends, family members, or spouses, mentor/mentee & Joint grant holders, personal relationship & Family members, doctoral thesis advisor/advisee, postdoctoral mentor/mentee \\
\hline
\multicolumn{1}{c}{} & Mentioned in the policy & Yes & Yes & No & Yes & Yes & Yes \\
\multicolumn{1}{c}{} & Only recent collaboration & Yes & Yes & N.A. & Yes & Yes & Yes \\
\multicolumn{1}{c}{} & Clear definintion of ``recent'' & 2 years & No & N.A. & 3 years & 5 years & 48 months \\
\multicolumn{1}{c}{\multirow{-4}{*}{\shortstack[l]{Editor-author\\associations due to\\recent collaboration}}} & Additional contraints & No & No & N.A. & No & No &  No \\
\hline
 & Mentioned in the policy & Yes & Yes & No & Yes & Yes & No \\
 & Same affiliation current & Yes & Yes & N.A. & Yes & Yes & N.A. \\
 \multirow{-3}{*}{\shortstack[l]{Editor-author\\associations due to\\same affiliation}}& Same affiliation in the past & No & Yes & N.A. & No & No & N.A. \\
\end{tabular}%
}
\caption{Same as Table~\ref{tab:policy} but for the six publishers analyzed in this study.}
\label{tab:publishers}
\end{table}

\clearpage
\begin{figure}[h]
\centering
\includegraphics[width=0.93\linewidth]{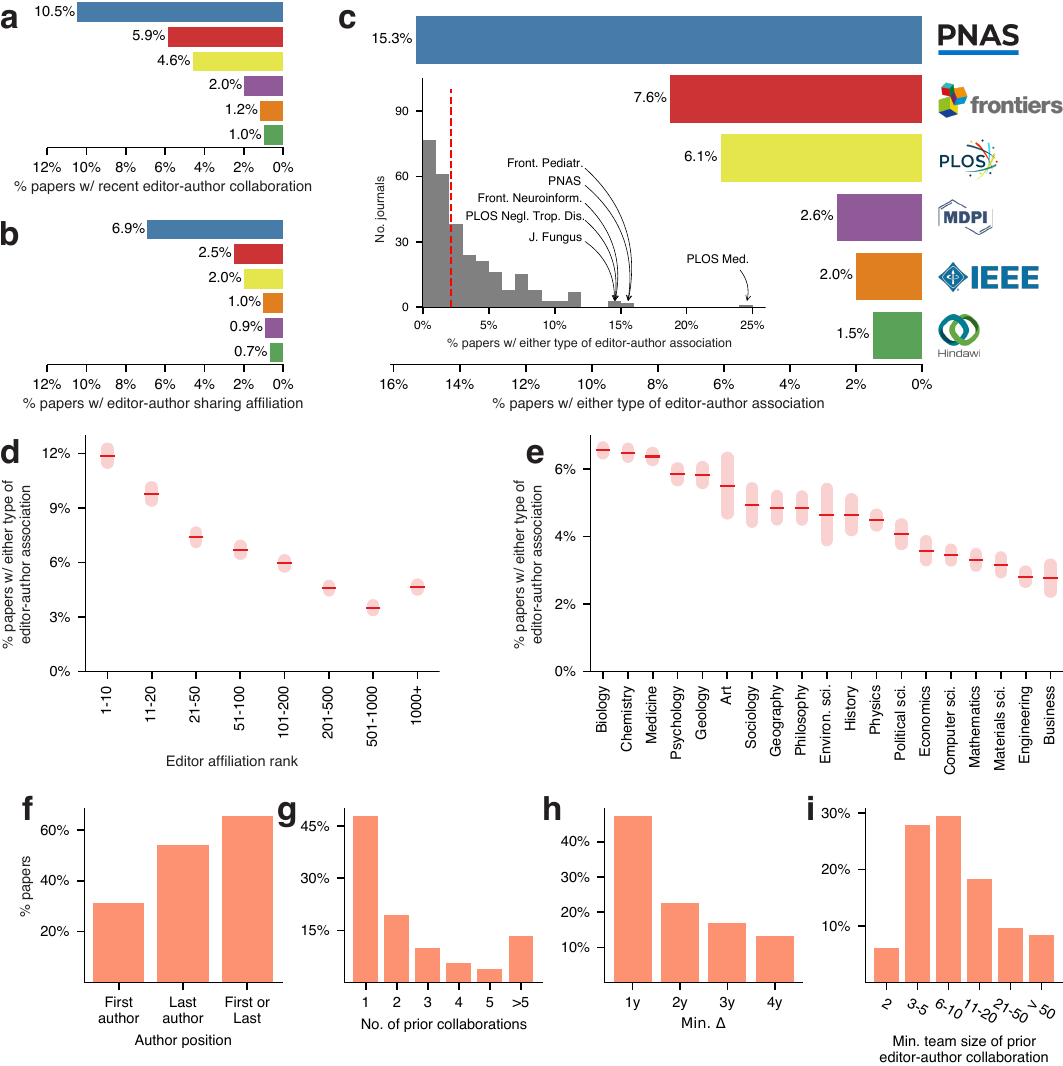}
\caption{\small\textbf{{Quantifying the percentage of papers with editor-author associations.}}
{\textbf{a}, In each publisher, the percentage of papers with recent editor-author collaboration(s).
\textbf{b}, In each publisher, the percentage of papers whose editor sharing the same affiliation with any author.
\textbf{c}, In each publisher, the percentage of papers with either type of editor-author association.} Inset shows the distribution of the {percentage} across all the journals in our dataset, highlighting the six journals with the highest {percentage}. The red dotted line highlights the median {percentage}.
{\textbf{d}, Percentage of papers with editor-author associations across editor's affiliation rank.}
\textbf{e}, {Percentage of papers with editor-author associations} across disciplines.
{
\textbf{f}, Among papers with editor-author associations, the percentage of papers where the association was introduced by the first author only, the last author only, or either first or last author.
\textbf{g}, Among papers with recent editor-author collaborations, the percentage of papers where the number of prior collaborations is between one and five, or greater than five.
\textbf{h}, Among papers with recent editor-author collaborations, the percentage of papers where the most recent collaboration happened one year (0-12 months), two years (13-24 months), three years (25-36 months), and four years (37-48 months) ago.
\textbf{i}, Among papers with recent editor-author collaborations, the percentage of papers where the minimum team size of prior collaboration falls in each of the six bins.
}
In ({\textbf{d}) and (\textbf{e}}), lines represent the mean {percentage} and the shaded regions represent 95\% confidence interval.
}
\label{coi_rate}
\end{figure}

\clearpage
\begin{SCfigure}[50][h]
\includegraphics[]{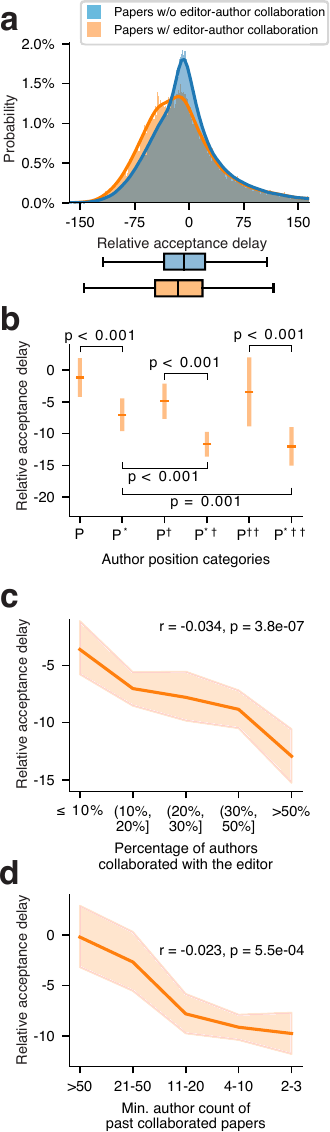}
\caption{\small{\textbf{Comparing the acceptance delay of papers with or without editor-author associations.}
\textbf{a}, Distributions of relative acceptance delay (RAD) of papers with and without {editor-author collaboration}. These distributions are summarized as boxplots, where boxes extend from the lower to upper quartile values, and whiskers extend until the 5th and the 95th percentiles. The lines represent the median.
{\textbf{b}, Papers are grouped based on the author position of the editor's collaborator in the focal paper as well as their positions in the prior collaboration. {The asterisk superscript ($p^*$) indicates that the handling editor of $p$ has collaborated with the first or last author of $p$. A dagger superscript ($p^{\dagger}$) indicates that an author of $p$ is the first or last author in a prior collaboration with the editor. Two daggers ($p^{\dagger\dagger}$) indicate that the editor and an author of $p$ were the first and last authors in a recent collaboration.}
The horizontal lines denote the mean RAD of each group of papers, and the vertical line denotes the 95\% confidence interval (CI). $p$-values are calculated using the Welch's t-test.}
{\textbf{c}}, Correlation between RAD and the percentage of authors that have recently collaborated with the editor.
{\textbf{d}}, Correlation between RAD and the minimum author count on any papers co-written by the editor and any authors of the focal paper in the past 48 months. 
In {(\textbf{c}) and (\textbf{d})}, lines represent the mean RAD and the shaded regions represent 95\% CI; the Pearson correlation coefficients ($r$) and the associated $p$-values are calculated using the original data (not binned).
}}
\label{rad}
\end{SCfigure}

\clearpage
\begin{figure}[h]
\centering
\includegraphics[width=\textwidth]{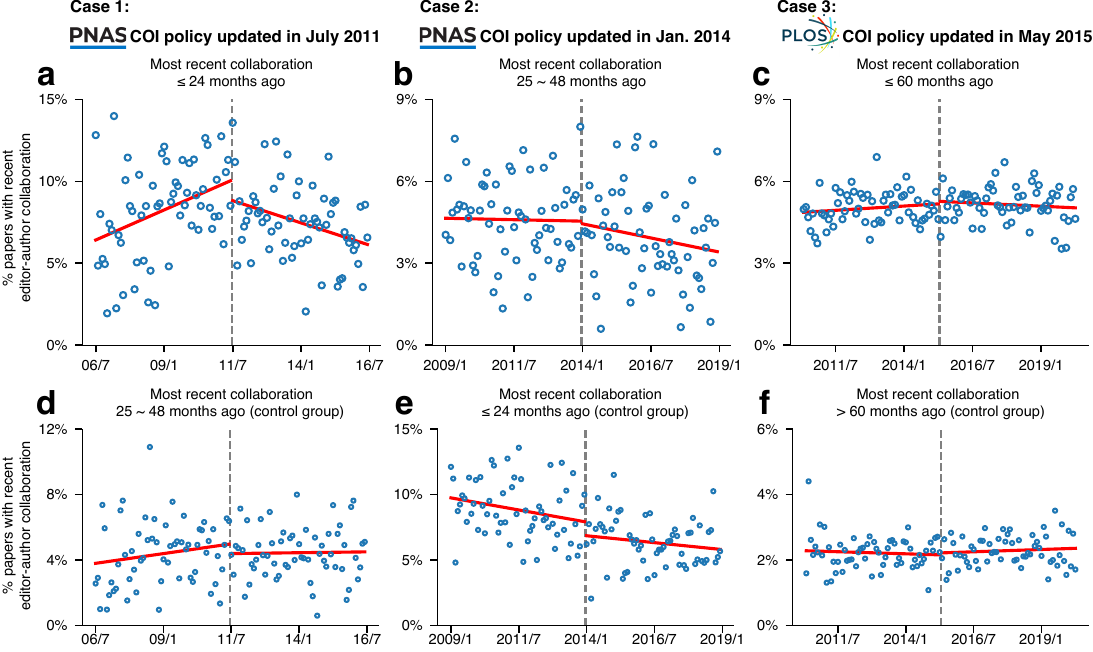}
\caption{\small{\textbf{Policies fail to eliminate papers with {recent editor-author collaboration}.}
The $x$-axis represent the submission date (in month) of papers. For any given month, the corresponding circle represents the percentage of papers submitted that month whose authors had a recent collaboration with its handling editor.
Recent collaboration is defined differently in each panel, as stated in their respective subtitles.
Dashed lines denote the time when a policy was introduced that prohibits editors from handling submissions by recent collaborators.
Red lines are fitted to the circles before and after the policy change using the OLS method.
The top row corresponds to the treatment groups, i.e., the papers that are targeted by the policy change, while the bottom row corresponds to the control groups, i.e., papers that are not affected by the change.
}}
\label{policy_change}
\end{figure}

\clearpage
\begin{SCfigure}[50][h]
\includegraphics[]{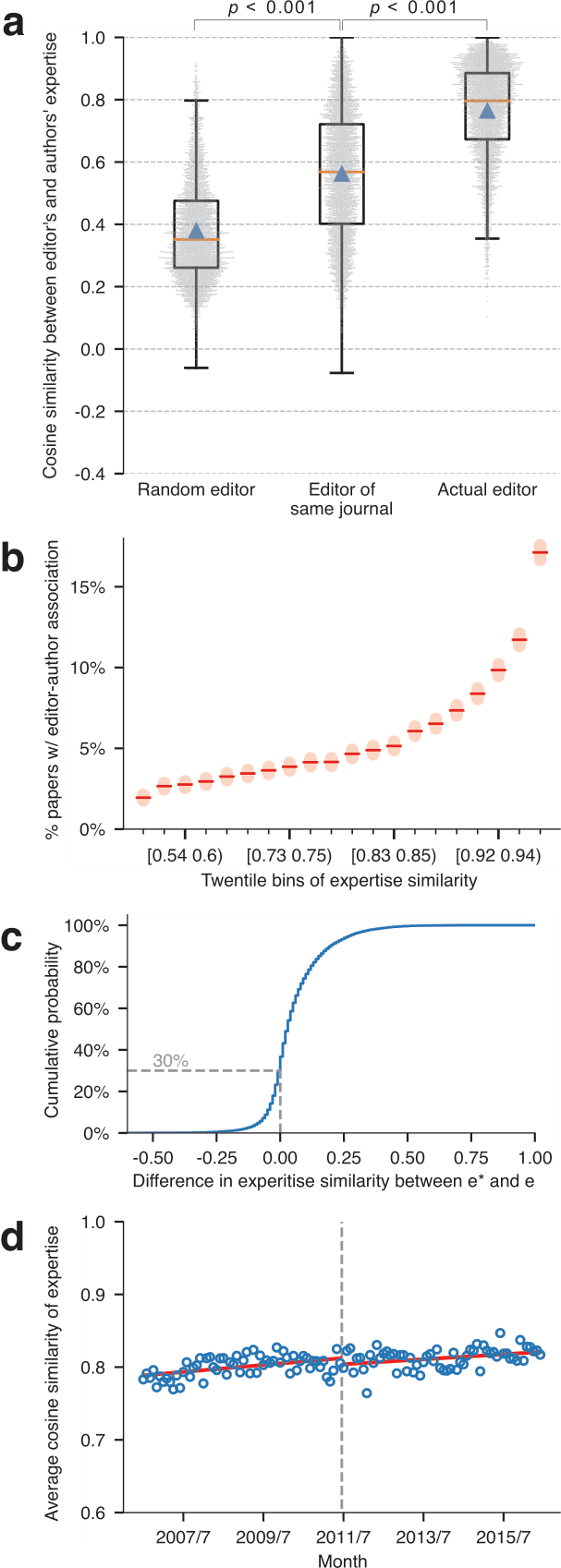} 
\caption{\small{\textbf{Expertise and {editor-author association}.}
\textbf{a}, Each data point correspond to a paper, and represents the cosine similarity between the paper and a given editor in terms of expertise. Here, the editor is either randomly sampled from our entire dataset (left), randomly sampled from the journal's editorial board (middle), or the actual editor who handled the paper (right). Boxes extend from the lower to upper quartile values, whiskers denote the interquartile range, lines denote the medians, and triangles denote the means. The swarmplots show the distribution of expertise similarity of 5000 randomly sampled papers. $p$-values are calculated using Mann–Whitney U test.
\textbf{b}, The average {percentage of papers with editor-author association} as a function of the expertise similarity between a paper and its handling editor. Here, expertise similarity is binned into 20 twentile bins. Lines represent the mean {percentage} while the shaded regions represent 95\%-CI.
\textbf{c}, The cumulative distribution of the expertise similarity between $e^*$ and $p$ minus that between $e$ and $p$, where $p$ is a paper handled by editor $e$ who has {an association with an author}, while $e^*$ is most relevant editorial board member who does not have {any association with any author}.
\textbf{d}, Temporal trend of the average expertise similarity between a paper and its handling editor in PNAS. For any given month, the corresponding circle represents the average expertise similarity between papers submitted in that month and their handling editors. Dashed line denotes July 2011 when PNAS updated its COI policy. Red lines are fitted to the circles before and after the policy change using the OLS method.
}}
\label{expertise}
\end{SCfigure}

\clearpage
\begin{figure}[h]
\centering
\includegraphics[]{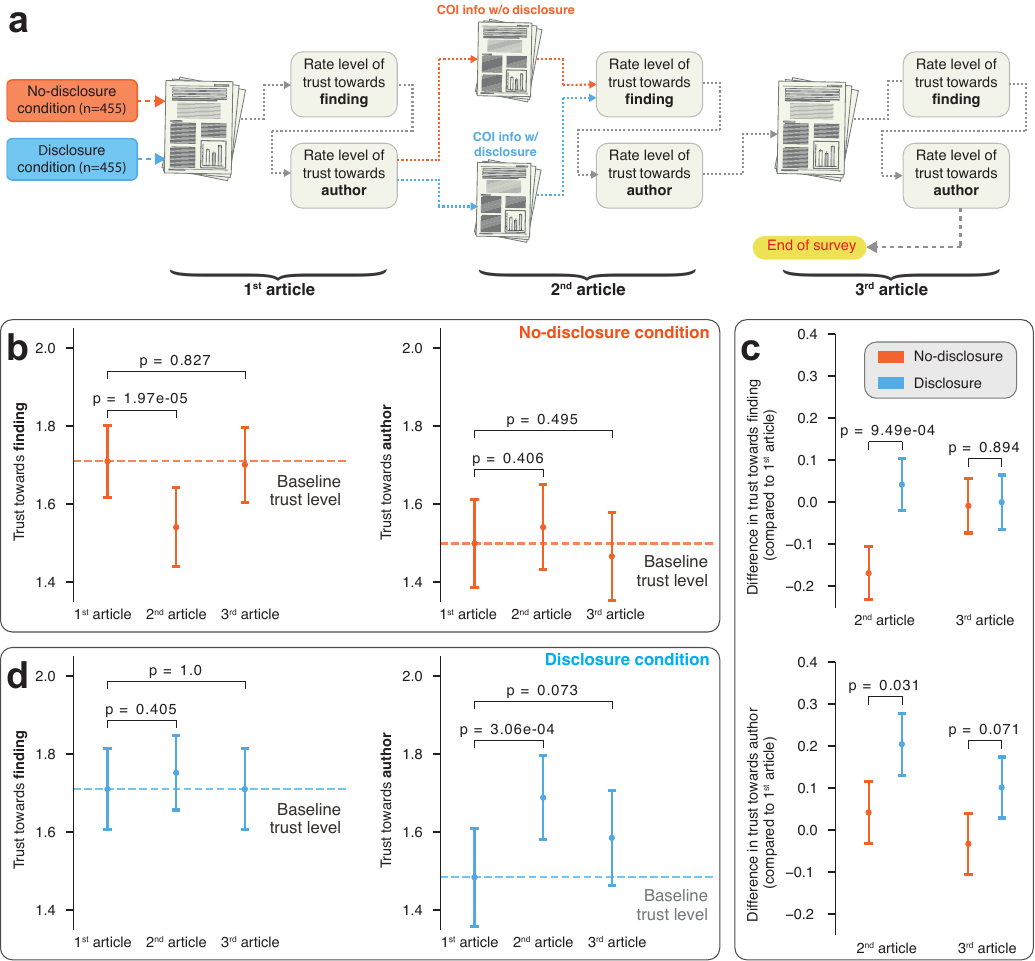}
\caption{\small{\textbf{Assessing the impact of editor-author associations and their disclosure on trust in the paper and the author.}}
\textbf{a}, Graphical representation of the experimental design and participant flow.
\textbf{b}, For the no-disclosure condition, the participants' level of trust towards the findings reported in, and the authors of, each article.
\textbf{c}, For both the disclosure condition (blue) and the no-disclosure condition (orange), the participants' level of trust towards the findings and authors of the second and third articles, relative to that of the first article.
\textbf{d}, The same as (\textbf{b}), but for the disclosure, rather than no-disclosure, condition. In {(\textbf{b})---(\textbf{d})}, dots represent the mean level of trust as reported by respondents while the error bars represent 95\%-CI.
}
\label{survey}
\end{figure}



\begin{thebibliography}{10}
\expandafter\ifx\csname url\endcsname\relax
  \def\url#1{\texttt{#1}}\fi
\expandafter\ifx\csname urlprefix\endcsname\relax\def\urlprefix{URL }\fi
\providecommand{\bibinfo}[2]{#2}
\providecommand{\eprint}[2][]{\url{#2}}

\bibitem{siler2015measuring}
\bibinfo{author}{Siler, K.}, \bibinfo{author}{Lee, K.} \&
  \bibinfo{author}{Bero, L.}
\newblock \bibinfo{title}{Measuring the effectiveness of scientific
  gatekeeping}.
\newblock \emph{\bibinfo{journal}{Proceedings of the National Academy of
  Sciences}} \textbf{\bibinfo{volume}{112}}, \bibinfo{pages}{360--365}
  (\bibinfo{year}{2015}).

\bibitem{CSE}
\bibinfo{title}{{CSE's White Paper on Promoting Integrity in Scientific Journal
  Publications}}.
\newblock
  \bibinfo{howpublished}{https://www.seaairweb.info/journal/3.CouncilofScientific-Editors-White-Paper.pdf,
  Council of Science Editors} (\bibinfo{year}{2012}).
\newblock \bibinfo{note}{Accessed on August 12, 2024}.

\bibitem{haivas2004editors}
\bibinfo{author}{Haivas, I.}, \bibinfo{author}{Schroter, S.},
  \bibinfo{author}{Waechter, F.} \& \bibinfo{author}{Smith, R.}
\newblock \bibinfo{title}{Editors' declaration of their own conflicts of
  interest}.
\newblock \emph{\bibinfo{journal}{{CMAJ}}} \textbf{\bibinfo{volume}{171}},
  \bibinfo{pages}{475--476} (\bibinfo{year}{2004}).

\bibitem{bosch2013financial}
\bibinfo{author}{Bosch, X.}, \bibinfo{author}{Pericas, J.~M.},
  \bibinfo{author}{Hernandez, C.} \& \bibinfo{author}{Doti, P.}
\newblock \bibinfo{title}{Financial, nonfinancial and editors' conflicts of
  interest in high-impact biomedical journals}.
\newblock \emph{\bibinfo{journal}{European journal of clinical investigation}}
  \textbf{\bibinfo{volume}{43}}, \bibinfo{pages}{660--667}
  (\bibinfo{year}{2013}).

\bibitem{faggion2021watching}
\bibinfo{author}{Faggion~Jr, C.~M.}
\newblock \bibinfo{title}{Watching the watchers: A report on the disclosure of
  potential conflicts of interest by editors and editorial board members of
  dental journals}.
\newblock \emph{\bibinfo{journal}{European Journal of Oral Sciences}}
  \textbf{\bibinfo{volume}{129}}, \bibinfo{pages}{e12823}
  (\bibinfo{year}{2021}).

\bibitem{smith2012accessibility}
\bibinfo{author}{Smith, E.}, \bibinfo{author}{Potvin, M.-J.} \&
  \bibinfo{author}{Williams-Jones, B.}
\newblock \bibinfo{title}{Accessibility and transparency of editor conflicts of
  interest policy instruments in medical journals}.
\newblock \emph{\bibinfo{journal}{Journal of Medical Ethics}}
  \textbf{\bibinfo{volume}{38}}, \bibinfo{pages}{679--684}
  (\bibinfo{year}{2012}).

\bibitem{plos2008making}
\bibinfo{title}{{PLOS Medicine Editors}, making sense of non-financial
  competing interests}.
\newblock \emph{\bibinfo{journal}{PLOS Medicine}} \textbf{\bibinfo{volume}{5}},
  \bibinfo{pages}{e199} (\bibinfo{year}{2008}).

\bibitem{COPE_COI}
\bibinfo{title}{Guidelines: Ethical guidelines for peer reviewers}.
\newblock \bibinfo{howpublished}{COPE (Committee on Publication Ethics),
  https://publicationethics.org/sites/default/files/ethical-guidelines-peer-reviewers-cope.pdf}.
\newblock \bibinfo{note}{Accessed on August 11, 2024}.

\bibitem{ICMJE}
\bibinfo{title}{{Disclosure of Financial and Non-Financial Relationships and
  Activities, and Conflicts of Interest}}.
\newblock \bibinfo{howpublished}{ICMJE,
  https://www.icmje.org/recommendations/browse/roles-and-responsibilities/author-responsibilities--conflicts-of-interest.html}.
\newblock \bibinfo{note}{Accessed on April 10, 2023}.

\bibitem{WAME2009COI}
\bibinfo{title}{Conflict of interest in peer-reviewed medical journals}.
\newblock \bibinfo{howpublished}{World Association of Mecial Editors}
  (\bibinfo{year}{2009}).
\newblock
  \urlprefix\url{https://www.wame.org/conflict-of-interest-in-peer-reviewed-medical-journals}.
\newblock \bibinfo{note}{Accessed on March 16, 2023}.

\bibitem{FrontiersPolicy}
\bibinfo{title}{Policies and publication ethics}.
\newblock \bibinfo{howpublished}{Frontiers,
  \url{https://www.frontiersin.org/guidelines/policies-and-publication-ethics}}.
\newblock \bibinfo{note}{Accessed on March 15, 2023}.

\bibitem{HindawiPolicy}
\bibinfo{title}{Publication ethics}.
\newblock \bibinfo{howpublished}{Hindawi,
  \url{https://www.hindawi.com/publish-research/authors/publication-ethics/}}.
\newblock \bibinfo{note}{Accessed on April 12, 2023}.

\bibitem{IEEEpolicy}
\bibinfo{title}{{IEEE} policies}.
\newblock \bibinfo{howpublished}{The Institute of Electrical and Electronics
  Engineers, Inc., New York, N.Y.} (\bibinfo{year}{2023}).
\newblock
  \urlprefix\url{https://www.ieee.org/content/dam/ieee-org/ieee/web/org/about/corporate/ieee-policies.pdf}.
\newblock \bibinfo{note}{Accessed on April 28, 2023}.

\bibitem{MDPIpolicy}
\bibinfo{title}{Research and publication ethics}.
\newblock \bibinfo{howpublished}{MDPI, \url{https://www.mdpi.com/ethics}}.
\newblock \bibinfo{note}{Accessed on April 25, 2023}.

\bibitem{PLOSpolicy}
\bibinfo{title}{Competing interests}.
\newblock \bibinfo{howpublished}{PLOS One,
  \url{https://journals.plos.org/plosone/s/competing-interests}}.
\newblock \bibinfo{note}{Accessed on April 12, 2023}.

\bibitem{PNASpolicy}
\bibinfo{title}{Editorial and journal policies}.
\newblock \bibinfo{howpublished}{PNAS Author Center,
  \url{https://www.pnas.org/author-center/editorial-and-journal-policies}}.
\newblock \bibinfo{note}{Accessed on April 26, 2023}.

\bibitem{IEEEpubPolicy}
\bibinfo{title}{{IEEE} publication services and products board operations
  manual 2023}.
\newblock \bibinfo{howpublished}{IEEE Publications, Piscataway, NJ}
  (\bibinfo{year}{2023}).
\newblock
  \urlprefix\url{https://pspb.ieee.org/images/files/PSPB/opsmanual.pdf}.
\newblock \bibinfo{note}{Accessed on April 28, 2023}.

\bibitem{science2023stripped}
\bibinfo{author}{Brainard, J.}
\newblock \bibinfo{title}{{Fast-growing open-access journals stripped of
  coveted impact factors}}.
\newblock
  \bibinfo{howpublished}{https://www.science.org/content/article/fast-growing-open-access-journals-stripped-coveted-impact-factors}
  (\bibinfo{year}{2023}).
\newblock \bibinfo{note}{Accessed on October 09, 2023}.

\bibitem{else2021scammers}
\bibinfo{author}{Else, H.}
\newblock \bibinfo{title}{Scammers impersonate guest editors to get sham papers
  published}.
\newblock \emph{\bibinfo{journal}{Nature}} \textbf{\bibinfo{volume}{599}},
  \bibinfo{pages}{361} (\bibinfo{year}{2021}).

\bibitem{mills2024special}
\bibinfo{author}{Mills, D.}, \bibinfo{author}{Mertkan, S.} \&
  \bibinfo{author}{Onurkan~Aliusta, G.}
\newblock \bibinfo{title}{‘special issue-ization’as a growth and revenue
  strategy: Reproduction by the “big five” and the risks for research
  integrity}.
\newblock \emph{\bibinfo{journal}{Accountability in Research}}
  \bibinfo{pages}{1--19} (\bibinfo{year}{2024}).

\bibitem{van2023more}
\bibinfo{author}{Van~Noorden, R.}
\newblock \bibinfo{title}{More than 10,000 research papers were retracted in
  2023---a new record}.
\newblock \emph{\bibinfo{journal}{Nature}} \textbf{\bibinfo{volume}{624}},
  \bibinfo{pages}{479--481} (\bibinfo{year}{2023}).

\bibitem{pfeffer2018queer}
\bibinfo{author}{Pfeffer, C.~A.}
\newblock \bibinfo{title}{Queer accounting: Methodological investments and
  disinvestments}.
\newblock \emph{\bibinfo{journal}{Other, please specify: Queer methods in
  sociology}} \bibinfo{pages}{304--325} (\bibinfo{year}{2018}).

\bibitem{else2021tortured}
\bibinfo{author}{Else, H.}
\newblock \bibinfo{title}{Tortured phrases’ give away fabricated}.
\newblock \emph{\bibinfo{journal}{Nature}} \textbf{\bibinfo{volume}{596}},
  \bibinfo{pages}{328--9} (\bibinfo{year}{2021}).

\bibitem{becher1994significance}
\bibinfo{author}{Becher, T.}
\newblock \bibinfo{title}{The significance of disciplinary differences}.
\newblock \emph{\bibinfo{journal}{Studies in Higher education}}
  \textbf{\bibinfo{volume}{19}}, \bibinfo{pages}{151--161}
  (\bibinfo{year}{1994}).

\bibitem{jaffe2014social}
\bibinfo{author}{Jaffe, K.}
\newblock \bibinfo{title}{Social and natural sciences differ in their research
  strategies, adapted to work for different knowledge landscapes}.
\newblock \emph{\bibinfo{journal}{PLoS One}} \textbf{\bibinfo{volume}{9}},
  \bibinfo{pages}{e113901} (\bibinfo{year}{2014}).

\bibitem{wray2022epistemic}
\bibinfo{author}{Wray, K.~B.}
\newblock \bibinfo{title}{The epistemic significance of collaborative
  research}.
\newblock \emph{\bibinfo{journal}{Philosophy of science}}
  \textbf{\bibinfo{volume}{69}}, \bibinfo{pages}{150--168}
  (\bibinfo{year}{2022}).

\bibitem{lariviere2006canadian}
\bibinfo{author}{Larivi{\`e}re, V.}, \bibinfo{author}{Gingras, Y.} \&
  \bibinfo{author}{Archambault, {\'E}.}
\newblock \bibinfo{title}{Canadian collaboration networks: A comparative
  analysis of the natural sciences, social sciences and the humanities}.
\newblock \emph{\bibinfo{journal}{Scientometrics}}
  \textbf{\bibinfo{volume}{68}}, \bibinfo{pages}{519--533}
  (\bibinfo{year}{2006}).

\bibitem{brogaard2014networks}
\bibinfo{author}{Brogaard, J.}, \bibinfo{author}{Engelberg, J.} \&
  \bibinfo{author}{Parsons, C.~A.}
\newblock \bibinfo{title}{Networks and productivity: Causal evidence from
  editor rotations}.
\newblock \emph{\bibinfo{journal}{Journal of Financial Economics}}
  \textbf{\bibinfo{volume}{111}}, \bibinfo{pages}{251--270}
  (\bibinfo{year}{2014}).

\bibitem{card2020editors}
\bibinfo{author}{Card, D.} \& \bibinfo{author}{DellaVigna, S.}
\newblock \bibinfo{title}{What do editors maximize? evidence from four
  economics journals}.
\newblock \emph{\bibinfo{journal}{Review of Economics and Statistics}}
  \textbf{\bibinfo{volume}{102}}, \bibinfo{pages}{195--217}
  (\bibinfo{year}{2020}).

\bibitem{sauermann2017authorship}
\bibinfo{author}{Sauermann, H.} \& \bibinfo{author}{Haeussler, C.}
\newblock \bibinfo{title}{Authorship and contribution disclosures}.
\newblock \emph{\bibinfo{journal}{Science Advances}}
  \textbf{\bibinfo{volume}{3}}, \bibinfo{pages}{e1700404}
  (\bibinfo{year}{2017}).

\bibitem{imbens2008regression}
\bibinfo{author}{Imbens, G.~W.} \& \bibinfo{author}{Lemieux, T.}
\newblock \bibinfo{title}{Regression discontinuity designs: A guide to
  practice}.
\newblock \emph{\bibinfo{journal}{Journal of Econometrics}}
  \textbf{\bibinfo{volume}{142}}, \bibinfo{pages}{615--635}
  (\bibinfo{year}{2008}).

\bibitem{anderson2014subways}
\bibinfo{author}{Anderson, M.~L.}
\newblock \bibinfo{title}{Subways, strikes, and slowdowns: The impacts of
  public transit on traffic congestion}.
\newblock \emph{\bibinfo{journal}{American Economic Review}}
  \textbf{\bibinfo{volume}{104}}, \bibinfo{pages}{2763--2796}
  (\bibinfo{year}{2014}).

\bibitem{reny2021opinion}
\bibinfo{author}{Reny, T.~T.} \& \bibinfo{author}{Newman, B.~J.}
\newblock \bibinfo{title}{The opinion-mobilizing effect of social protest
  against police violence: Evidence from the 2020 george floyd protests}.
\newblock \emph{\bibinfo{journal}{American Political Science Review}}
  \textbf{\bibinfo{volume}{115}}, \bibinfo{pages}{1499--1507}
  (\bibinfo{year}{2021}).

\bibitem{retraction_watch}
\bibinfo{title}{{Retraction Watch}}.
\newblock \urlprefix\url{https://retractionwatch.com/}.

\bibitem{hausman2018regression}
\bibinfo{author}{Hausman, C.} \& \bibinfo{author}{Rapson, D.~S.}
\newblock \bibinfo{title}{Regression discontinuity in time: Considerations for
  empirical applications}.
\newblock \emph{\bibinfo{journal}{Annual Review of Resource Economics}}
  \textbf{\bibinfo{volume}{10}}, \bibinfo{pages}{533--552}
  (\bibinfo{year}{2018}).

\bibitem{resnik2016ensuring}
\bibinfo{author}{Resnik, D.~B.} \& \bibinfo{author}{Elmore, S.~A.}
\newblock \bibinfo{title}{Ensuring the quality, fairness, and integrity of
  journal peer review: A possible role of editors}.
\newblock \emph{\bibinfo{journal}{Science and Engineering Ethics}}
  \textbf{\bibinfo{volume}{22}}, \bibinfo{pages}{169--188}
  (\bibinfo{year}{2016}).

\bibitem{PLOSPeerReviewPolicy}
\bibinfo{title}{Editorial and peer review process}.
\newblock
  \urlprefix\url{https://journals.plos.org/plosone/s/editorial-and-peer-review-process}.
\newblock \bibinfo{note}{Accessed on December 4, 2023}.

\bibitem{HindawiPeerReviewPolicy}
\bibinfo{title}{Editors}.
\newblock \urlprefix\url{https://www.hindawi.com/publish-research/editors/}.
\newblock \bibinfo{note}{Accessed on December 4, 2023}.

\bibitem{resnik2018conflict}
\bibinfo{author}{Resnik, D.~B.} \& \bibinfo{author}{Elmore, S.~A.}
\newblock \bibinfo{title}{Conflict of interest in journal peer review}.
\newblock \emph{\bibinfo{journal}{Toxicologic Pathology}}
  \textbf{\bibinfo{volume}{46}}, \bibinfo{pages}{112--114}
  (\bibinfo{year}{2018}).

\bibitem{gottlieb2017should}
\bibinfo{author}{Gottlieb, J.~D.} \& \bibinfo{author}{Bressler, N.~M.}
\newblock \bibinfo{title}{How should journals handle the conflict of interest
  of their editors?: who watches the “watchers”?}
\newblock \emph{\bibinfo{journal}{Jama}} \textbf{\bibinfo{volume}{317}},
  \bibinfo{pages}{1757--1758} (\bibinfo{year}{2017}).

\bibitem{friedman2002impact}
\bibinfo{author}{Friedman, P.~J.}
\newblock \bibinfo{title}{The impact of conflict of interest on trust in
  science}.
\newblock \emph{\bibinfo{journal}{Science and Engineering ethics}}
  \textbf{\bibinfo{volume}{8}}, \bibinfo{pages}{413--420}
  (\bibinfo{year}{2002}).

\bibitem{malivcki2021systematic}
\bibinfo{author}{Mali{\v{c}}ki, M.}, \bibinfo{author}{Jeron{\v{c}}i{\'c}, A.},
  \bibinfo{author}{Aalbersberg, I.~J.}, \bibinfo{author}{Bouter, L.} \&
  \bibinfo{author}{Ter~Riet, G.}
\newblock \bibinfo{title}{Systematic review and meta-analyses of studies
  analysing instructions to authors from 1987 to 2017}.
\newblock \emph{\bibinfo{journal}{Nature Communications}}
  \textbf{\bibinfo{volume}{12}}, \bibinfo{pages}{5840} (\bibinfo{year}{2021}).

\bibitem{jpubhelthpolicy}
\bibinfo{title}{Conflict of interest policy}.
\newblock \bibinfo{howpublished}{Public Health Nutrition, Cambridge Core}.
\newblock
  \urlprefix\url{https://www.cambridge.org/core/journals/public-health-nutrition/information/conflict-of-interest-policy}.
\newblock \bibinfo{note}{Accessed on December 17, 2024}.

\bibitem{nicholas2023never}
\bibinfo{author}{Nicholas, D.} \emph{et~al.}
\newblock \bibinfo{title}{Never mind predatory publishers… what about
  ‘grey’publishers?}
\newblock \emph{\bibinfo{journal}{Profesional de la informaci{\'o}n}}
  \textbf{\bibinfo{volume}{32}} (\bibinfo{year}{2023}).

\bibitem{kean2009pnas}
\bibinfo{author}{Kean, S.}
\newblock \bibinfo{title}{{PNAS} nixes special privileges for (most) papers}
  (\bibinfo{year}{2009}).

\bibitem{nishikawa2022100}
\bibinfo{author}{Nishikawa-Pacher, A.}
\newblock \bibinfo{title}{Who are the 100 largest scientific publishers by
  journal count? a webscraping approach}.
\newblock \emph{\bibinfo{journal}{Journal of Documentation}}
  \textbf{\bibinfo{volume}{78}}, \bibinfo{pages}{450--463}
  (\bibinfo{year}{2022}).

\bibitem{retraction2022PNAS}
\bibinfo{author}{Oransky, I.}
\newblock \bibinfo{title}{{White House} official banned from publishing in
  {PNAS} following retraction}.
\newblock \bibinfo{howpublished}{{Retraction Watch}} (\bibinfo{year}{2022}).
\newblock \bibinfo{note}{Accessed on March 16, 2023}.

\bibitem{bero2016having}
\bibinfo{author}{Bero, L.~A.} \& \bibinfo{author}{Grundy, Q.}
\newblock \bibinfo{title}{Why having a (nonfinancial) interest is not a
  conflict of interest}.
\newblock \emph{\bibinfo{journal}{PLOS Biology}} \textbf{\bibinfo{volume}{14}},
  \bibinfo{pages}{e2001221} (\bibinfo{year}{2016}).

\bibitem{coles2022build}
\bibinfo{author}{Coles, N.~A.}, \bibinfo{author}{Hamlin, J.~K.},
  \bibinfo{author}{Sullivan, L.~L.}, \bibinfo{author}{Parker, T.~H.} \&
  \bibinfo{author}{Altschul, D.}
\newblock \bibinfo{title}{Build up big-team science}.
\newblock \emph{\bibinfo{journal}{Nature}} \textbf{\bibinfo{volume}{601}},
  \bibinfo{pages}{505--507} (\bibinfo{year}{2022}).

\bibitem{katz1997research}
\bibinfo{author}{Katz, J.~S.} \& \bibinfo{author}{Martin, B.~R.}
\newblock \bibinfo{title}{What is research collaboration?}
\newblock \emph{\bibinfo{journal}{Research Policy}}
  \textbf{\bibinfo{volume}{26}}, \bibinfo{pages}{1--18} (\bibinfo{year}{1997}).

\bibitem{thelwall2020large}
\bibinfo{author}{Thelwall, M.}
\newblock \bibinfo{title}{Large publishing consortia produce higher citation
  impact research but coauthor contributions are hard to evaluate}.
\newblock \emph{\bibinfo{journal}{Quantitative Science Studies}}
  \textbf{\bibinfo{volume}{1}}, \bibinfo{pages}{290--302}
  (\bibinfo{year}{2020}).

\bibitem{authorship2008}
\bibinfo{title}{Authorship matters}.
\newblock \emph{\bibinfo{journal}{Nature Mater}} \textbf{\bibinfo{volume}{7}}
  (\bibinfo{year}{2008}).

\bibitem{maddi2024beyond}
\bibinfo{author}{Maddi, A.} \& \bibinfo{author}{da~Silva, J. A.~T.}
\newblock \bibinfo{title}{Beyond authorship: Analyzing contributions in plos
  one and the challenges of appropriate attribution}.
\newblock \emph{\bibinfo{journal}{Journal of Data and Information Science}}
  \textbf{\bibinfo{volume}{9}}, \bibinfo{pages}{88--115}
  (\bibinfo{year}{2024}).

\bibitem{smith1958trend}
\bibinfo{author}{Smith, M.}
\newblock \bibinfo{title}{The trend toward multiple authorship in psychology.}
\newblock \emph{\bibinfo{journal}{American psychologist}}
  \textbf{\bibinfo{volume}{13}}, \bibinfo{pages}{596} (\bibinfo{year}{1958}).

\bibitem{melin1996studying}
\bibinfo{author}{Melin, G.} \& \bibinfo{author}{Persson, O.}
\newblock \bibinfo{title}{Studying research collaboration using
  co-authorships}.
\newblock \emph{\bibinfo{journal}{Scientometrics}}
  \textbf{\bibinfo{volume}{36}}, \bibinfo{pages}{363--377}
  (\bibinfo{year}{1996}).

\bibitem{wagner2005network}
\bibinfo{author}{Wagner, C.~S.} \& \bibinfo{author}{Leydesdorff, L.}
\newblock \bibinfo{title}{Network structure, self-organization, and the growth
  of international collaboration in science}.
\newblock \emph{\bibinfo{journal}{Research Policy}}
  \textbf{\bibinfo{volume}{34}}, \bibinfo{pages}{1608--1618}
  (\bibinfo{year}{2005}).

\bibitem{lerner2023micro}
\bibinfo{author}{Lerner, J.} \& \bibinfo{author}{H{\^a}ncean, M.-G.}
\newblock \bibinfo{title}{Micro-level network dynamics of scientific
  collaboration and impact: Relational hyperevent models for the analysis of
  coauthor networks}.
\newblock \emph{\bibinfo{journal}{Network Science}}
  \textbf{\bibinfo{volume}{11}}, \bibinfo{pages}{5--35} (\bibinfo{year}{2023}).

\bibitem{luborsky1999researcher}
\bibinfo{author}{Luborsky, L.} \emph{et~al.}
\newblock \bibinfo{title}{The researcher's own therapy allegiances: A" wild
  card" in comparisons of treatment efficacy.}
\newblock \emph{\bibinfo{journal}{Clinical psychology: Science and practice}}
  \textbf{\bibinfo{volume}{6}}, \bibinfo{pages}{95} (\bibinfo{year}{1999}).

\bibitem{MDPIPeerReviewPolicy}
\bibinfo{title}{{The MDPI Editorial Process}}.
\newblock \urlprefix\url{https://www.mdpi.com/editorial_process}.
\newblock \bibinfo{note}{Accessed on December 4, 2023}.

\bibitem{taheri2021discrepancies}
\bibinfo{author}{Taheri, C.} \emph{et~al.}
\newblock \bibinfo{title}{Discrepancies in self-reported financial conflicts of
  interest disclosures by physicians: a systematic review}.
\newblock \emph{\bibinfo{journal}{BMJ Open}} \textbf{\bibinfo{volume}{11}},
  \bibinfo{pages}{e045306} (\bibinfo{year}{2021}).

\bibitem{maeyama2014palindromic}
\bibinfo{author}{Maeyama, J.-i.} \emph{et~al.}
\newblock \bibinfo{title}{A palindromic cpg-containing phosphodiester
  oligodeoxynucleotide as a mucosal adjuvant stimulates plasmacytoid dendritic
  cell-mediated th1 immunity}.
\newblock \emph{\bibinfo{journal}{PLOS One}} \textbf{\bibinfo{volume}{9}},
  \bibinfo{pages}{e88846} (\bibinfo{year}{2014}).

\bibitem{liu2023non}
\bibinfo{author}{Liu, F.}, \bibinfo{author}{Rahwan, T.} \&
  \bibinfo{author}{AlShebli, B.}
\newblock \bibinfo{title}{Non-white scientists appear on fewer editorial
  boards, spend more time under review, and receive fewer citations}.
\newblock \emph{\bibinfo{journal}{Proceedings of the National Academy of
  Sciences}} \textbf{\bibinfo{volume}{120}}, \bibinfo{pages}{e2215324120}
  (\bibinfo{year}{2023}).

\bibitem{maydefinition}
\bibinfo{title}{May not definition}.
\newblock \bibinfo{howpublished}{Law Insider}.
\newblock \urlprefix\url{https://www.lawinsider.com/dictionary/may-not}.
\newblock \bibinfo{note}{Accessed on July 16, 2024}.

\bibitem{shoulddefinition}
\bibinfo{title}{Should not definition}.
\newblock \bibinfo{howpublished}{Law Insider}.
\newblock \urlprefix\url{https://www.lawinsider.com/dictionary/should-not}.
\newblock \bibinfo{note}{Accessed on July 16, 2024}.

\bibitem{wayback}
\bibinfo{title}{Internet archive: Wayback machine}.
\newblock \bibinfo{howpublished}{https://archive.org/web/}
  (\bibinfo{year}{2023}).

\bibitem{sinha2015overview}
\bibinfo{author}{Sinha, A.} \emph{et~al.}
\newblock \bibinfo{title}{An overview of {M}icrosoft {A}cademic {S}ervice
  ({MAS}) and applications}.
\newblock In \emph{\bibinfo{booktitle}{Proceedings of the 24th international
  conference on {World Wide Web}}}, \bibinfo{pages}{243--246}
  (\bibinfo{year}{2015}).

\bibitem{wang2019review}
\bibinfo{author}{Wang, K.} \emph{et~al.}
\newblock \bibinfo{title}{A review of {M}icrosoft {A}cademic {S}ervices for
  science of science studies}.
\newblock \emph{\bibinfo{journal}{Front. Big Data}}
  \textbf{\bibinfo{volume}{2}}, \bibinfo{pages}{45} (\bibinfo{year}{2019}).

\bibitem{wang2020microsoft}
\bibinfo{author}{Wang, K.} \emph{et~al.}
\newblock \bibinfo{title}{Microsoft academic graph: When experts are not
  enough}.
\newblock \emph{\bibinfo{journal}{Quantitative Science Studies}}
  \textbf{\bibinfo{volume}{1}}, \bibinfo{pages}{396--413}
  (\bibinfo{year}{2020}).

\bibitem{shanghairanking}
\bibinfo{title}{{ShanghaiRanking}}.
\newblock \bibinfo{howpublished}{https://www.shanghairanking.com/}
  (\bibinfo{year}{2023}).
\newblock \bibinfo{note}{Accessed October 22, 2023}.

\bibitem{shanghairankingmethod}
\bibinfo{title}{{ShanghaiRanking's Academic Ranking of World Universities
  Methodology 2023}}.
\newblock
  \bibinfo{howpublished}{https://www.shanghairanking.com/methodology/arwu/2023}
  (\bibinfo{year}{2023}).
\newblock \bibinfo{note}{Accessed on October 22, 2023}.

\bibitem{radun2023nonfinancial}
\bibinfo{author}{Radun, I.}
\newblock \bibinfo{title}{Nonfinancial conflict of interest in peer-review:
  Some notes for discussion}.
\newblock \emph{\bibinfo{journal}{Accountability in Research}}
  \textbf{\bibinfo{volume}{30}}, \bibinfo{pages}{331--342}
  (\bibinfo{year}{2023}).

\bibitem{constantino2023representing}
\bibinfo{author}{Constantino, I.}, \bibinfo{author}{Kojaku, S.},
  \bibinfo{author}{Fortunato, S.} \& \bibinfo{author}{Ahn, Y.-Y.}
\newblock \bibinfo{title}{Representing the disciplinary structure of physics: A
  comparative evaluation of graph and text embedding methods}.
\newblock \emph{\bibinfo{journal}{arXiv preprint arXiv:2308.15706}}
  (\bibinfo{year}{2023}).

\bibitem{cappelletti2023grape}
\bibinfo{author}{Cappelletti, L.} \emph{et~al.}
\newblock \bibinfo{title}{Grape for fast and scalable graph processing and
  random-walk-based embedding}.
\newblock \emph{\bibinfo{journal}{Nature Computational Science}}
  \textbf{\bibinfo{volume}{3}}, \bibinfo{pages}{552--568}
  (\bibinfo{year}{2023}).

\bibitem{COPE_editor_COI}
\bibinfo{title}{Conflicts of interest between authors and editors}.
\newblock \bibinfo{howpublished}{COPE (Committee on Publication Ethics),
  https://publicationethics.org/guidance/case/conflicts-interest-between-authors-and-editors}.
\newblock \bibinfo{note}{Accessed on August 11, 2024}.

\end{thebibliography}
\end{document}